%% file: Prophylaxy.tex
\documentclass [11pt]{article}
\usepackage[pdftex]{graphicx}
\usepackage{graphicx}
\usepackage{float}
\usepackage{bm}
\usepackage{color}
\usepackage{hyperref}
\usepackage[normalem]{ulem}
\hypersetup{colorlinks=true,
            linkcolor={green},
            citecolor={blue}}
\usepackage[makeroom]{cancel}
\usepackage{amsmath, amsfonts, latexsym}
\usepackage{paralist}
\usepackage{mathabx}
\usepackage{mathrsfs,amssymb}
\usepackage{version}
\usepackage[round]{natbib}
\usepackage{bbm}
\usepackage{flafter}
\usepackage{ulem}
\usepackage{dsfont}
\usepackage{amsthm}
\usepackage[english]{babel}
\usepackage{cancel}
\usepackage{nicefrac}

\numberwithin{equation}{section}
\renewcommand{\theequation}{\arabic{section}.\arabic{equation}}

\input{basecom}

\newcommand{\V}{\mathbb V}
\newcommand{\p}{\mathbb P}
\newcommand{\N}{\mathbb N}

\newcommand{\defeq}{\mathrel{\mathop:}=}

\def\bdelta{\bm{\delta}}

\def\x{\textbf{\textit{x}}}
\def\e{\textbf{\textit{e}}}
\def\u{\textbf{\textit{u}}}
\def\v{\textbf{\textit{v}}}
\def\V{\textbf{\textit{V}}}

\def\y{\textbf{\textit{y}}}
\def\1b{\textbf{{1}}}

\DeclareMathOperator{\diago}{diag}

\DeclareMathOperator{\POAL}{PoA^{Loc}}
\DeclareMathOperator{\POAG}{PoA^{Glo}}
\DeclareMathOperator{\POK}{PoK}

 \theoremstyle{plain}
 \newtheorem{theo}{Theorem}[section]
 \newtheorem{lem}{Lemma}[section]
 \newtheorem{prop}{Proposition}[section]

 \newtheorem{dfn}{Definition}[section]
 \newtheorem{rem}{Remark}[section]

 \theoremstyle{definition}
  
 \newtheorem{ex}{Example}[section]

 \theoremstyle{remark}
 
 \newtheorem{assump}{Assumption}[subsection]

\newcommand{\yellow}[1]{}

\newcommand{\fact}[1]{#1\mathpunct{}!}

\usepackage{titling}

\thanksmarkseries{alph}

\newenvironment{itemize*}
 {\begin{itemize}
    \setlength{\itemsep}{0pt}
    \setlength{\parskip}{0pt}
    \setlength{\parsep}{0pt}}
  {\end{itemize}}

\usepackage{paralist}
\usepackage{enumitem}

 \title{ \LARGE Prophylaxis of Epidemic Spreading with Transient Dynamics}
 \author{G\'eraldine Bouveret\thanks{Nanyang Technological University, School of Physical and Mathematical Sciences, 21 Nanyang Link, 637371
Singapore, Email: \sf{geraldine.bouveret@ntu.edu.sg}} \and Antoine Mandel \thanks{Universit\'e Paris 1 Panth\'eon-Sorbonne, 17 rue de la Sorbonne, Paris, IL 75005, France, Email: \sf antoine.mandel@univ-paris1.fr}
}

\begin{document}

 \maketitle
\vspace{-0.5cm}
 \begin{abstract}
 We investigate the containment of epidemic spreading in networks  from a normative point of view. We consider a susceptible/infected model in which agents can invest  in order to reduce the contagiousness of network links. In this setting, we study the relationships between social efficiency, individual behaviours and network structure. First, we exhibit an upper bound on  the Price of Anarchy and prove that the level of inefficiency can scale up to linearly with the number of agents.
Second,  we prove that policies of uniform reduction of interactions satisfy some optimality conditions in a vast range of networks. In setting where no central authority can enforce such stringent policies, we consider as a type of second-best policy the shift from a local to a global game by allowing agents to subsidise investments in contagiousness reduction in the global rather than in the local network. We then characterise the scope for Pareto improvement opened by such policies through a notion of Price of Autarky, measuring the ratio between social welfare at a global and a local equilibrium.
Overall, our results show that individual behaviours can be extremely inefficient in the face of epidemic propagation but that policy can take advantage of the network structure to design efficient containment policies.
    \end{abstract}


\vspace{0.3cm}

\section{Introduction}
Limiting epidemic spreading is today's key policy concern in the context of the COVID-19 pandemic. Beyond diseases, a number of socio-economic bads diffuse through social  networks via epidemic-like processes, e.g., financial distress or fake news. However, there is, to our knowledge, no normative analysis of the challenges posed by the containment of epidemic spreading. This is the issue we address in this paper.

The containment of epidemic processes defines a specific class of externality problems: through prophylactic investment, agents can reduce their own contamination risk but also reduce the risk of contagion of their peers in the network. The external effect hence created has certain features of a public good as the investment of each agent benefits to all the agents to whom it is connected. However, the magnitude of the effect depends on the specific connectivity between each pair of agents and thus on the structure of the network. In this setting, we aim to characterise, as a function of the network structure, (i) socially optimal containment policies, (ii) inefficiencies induced by individual strategic behaviours, and  (iii) the type of measures that can be implemented  to overcome these inefficiencies.

We place ourselves in a setting where the network structure is given, each agent can be initially contaminated with a certain probability, and contagion spreads through network links proportionally to their contagiousness. Once infected, agents remain so permanently, i.e. we consider a susceptible/infected type of  model according to the epidemiological terminology. In this context, agents aim at minimising their probability of contagion before a given date. In  a narrow interpretation, this date can be seen as the expected date at which a treatment will be available.  In a broader sense, the objective of each individual is to reduce the speed of incoming epidemic propagation. We assume that agents can invest in the network to reduce the speed of contagion. More precisely, they can decrease the contagiousness of links at a fixed linear cost. As the impact of individual investments depends on global contagiousness, and hence on the investment of other players, the situation defines a non-cooperative game. We consider two variants of the game. The local game in which an agent can only invest in the links through which it is connected. The global game in which an agent can invest in each link of the network.
The local game naturally applies to settings where agents are individuals that can take individual and costly measures to limit their social interactions. The global game corresponds to a more complex setting where agents are usually organisations (regions, countries) that are involved in a scheme that allows one agent to subsidise, directly or indirectly, the investment of other agents in the reduction of contagiousness.

Our main results characterise the relationships between social efficiency, individual behaviours and network structure. First, we derive an upper bound on  the Price of Anarchy (PoA). We show that in  worst cases the level of inefficiency can scale up to linearly with the number of agents. This strongly calls for public policy interventions. In this respect, we study the optimality of a policy of uniform reduction of interactions, akin to social distancing measures put in place during the COVID-19 pandemic, in a wide range of networks. This latter result provides normative foundations for the social distancing policies implemented during the COVID-19 pandemic.  The implementation of such policies nevertheless requires the existence of an authority with sufficient legitimacy to implement such coercive measures. It can be thus implemented in a domestic context but is much harder to implement at the global scale, unless all agents/countries have individual incentives to do so. If this is not the case, we regard the shift from a local to a global game as a type of second-best policy. In the latter game, agents can subsidise investments towards contagiousness diminution in the global rather than in the local network. The scope for Pareto improvement generated by such policies is then characterised through a notion of Price of Autarky (PoK), which assesses the ratio between social welfare at a global and a local equilibrium. We derive a lower bound on this PoK as a function of the network structure and thus give sufficient conditions under which a shift to the global game actually induces a Pareto improvement.
Overall, the results derived not only underline the possible extreme inefficiency of individual behaviours to limit epidemic propagation, but also the potential for policies to benefit from the network structure to design efficient containment policies.

The remaining of this paper is organised as follows. Section \ref{se lit} reviews the related literature. Section \ref{se model} introduces epidemic dynamics as well as our behavioural model of the containment of epidemic spreading. Section \ref{se eqsoc} provides our main results on the relationship between individual behaviours, social efficiency and network structure. Section \ref{se pok} introduces the notion of PoK and applies it to our setting. Section \ref{se conc} concludes. An appendix with the proofs of the main results is provided.

 \section{Related literature}\label{se lit}

 The paper builds on the very large literature on the optimal design and defense of networks (see, e.g., \cite{bravard17}) and on epidemic spreading in networks. The latter literature has been extensively reviewed in \citet{pastor2015epidemic} and generally  combines an epidemiological model with a diffusion model. The epidemiological model describes the characteristics of the disease via the set of states each agent can assume, e.g., susceptible/infected (SI), susceptible/infected/susceptible (SIS),  susceptible/infected/removed (SIR), and the probabilities of transition between these.  The diffusion model  considers that the set of agents is embedded in a network structure through which the disease spreads in a stochastic manner. Overall, the micro-level epidemic diffusion model is a  continuous-time Markov chain model whose state space corresponds to the complete epidemiological status of the population. This state space is however too large for the full model to be computationally or analytically tractable. A large strand of the literature has thus focused on the development of good approximations of the dynamics, see, e.g., \cite{chakrabarti2008epidemic,draief2006thresholds,ganesh2005effect,mei2017dynamics,prakash2012threshold,ruhi2016analysis,van2009virus,wang2003epidemic}.  To the best of our knowledge, the most precise approximation of the dynamics in the SIS/SIR setting is the $N-$ intertwined model of  \cite{van2009virus}. This model uses one (mean-field) approximation in the exact SIS model to convert the exact model into a set of $N$ non-linear differential equations. This transformation allows analytic computations that remain impossible with other more precise SIS models and renders the model relevant for any arbitrary graph. The $N-$ intertwined model upper bounds the exact model for finite networks of size $N$ and its accuracy improves with $N$.  \cite{inh2008} have  extended the  model to the heterogeneous case where the infection and curing rates depend on the node. Later, \cite{van2013decay} has analytically derived the decay rate of SIS epidemics on a complete graph, while \cite{van2014exact} has proposed an exact Markovian SIS and SIR epidemics on networks together with an upper bound for the epidemic threshold.

Most of this literature has focused on SIS/SIR models in which there exists an epidemic threshold above which the disease spreads exponentially.  A key concern has thus been the approximation of the epidemic threshold as a function of the characteristics of the network, and subsequently the determination of immunisation policies that allow to  reach the\textit{ below-the-threshold} regime (see, e.g., \cite{chen2016eigen,chen2015node,holme2002attack,preciado2013optimal,preciado2014optimal,saha2015approximation,schneider2011suppressing,van2011decreasing}).

A handful of studies has adopted a normative approach to the issue using a game-theoretic setting.
\cite{omic2009protecting} consider a $N-$intertwined SIS epidemic model, in which agents can invest in their curing rate. They prove the existence of a Nash Equilibrium and derive its characteristics as a function of the network structure. They provide a measure of social efficiency through the PoA. They  also investigate two types of policies to reduce contagiousness. The first one plays with the influence of the relative prices of protection while the second one relies on the enforcement of an upper bound on infection probabilities. \cite{hayel2014complete} have  also analysed decentralised optimal protection strategies in a SIS epidemic model. However, in their case, the curing and infection rates are fixed and each node can either invest in an antivirus to be fully protected or invest in a recovery software once infected. They  show that the game is a potential one, expressed the pure Nash Equilibrium for a single community/fully-mesh network in a closed form, and establish the existence and uniqueness of a mixed Nash Equilibrium. They  also provide a characterisation of the PoA.  Finally, \cite{goyal2015interaction}  examine, in a two-period model, the trade-off faced by individuals between reducing interaction and buying protection, and its impacts on infection rates. They analyse the equilibrium levels of interaction and protection as well as the infection rate of the population, and show the existence of a unique equilibrium. They highlight that individuals investing in protection are more willing to interact than those who do not invest, and establish the non-monotonic effects of changes in the contagiousness of a disease.

Yet, most of these contributions focus on situations where (i) some form of vaccine or treatment is available and (ii) dynamics are of the SIS/SIR type.  Our attention is rather on situations where there is no known cure to the epidemic and where the objective is to delay its propagation through investments in the reduction of contagiousness. Therefore, we focus on the transient dynamics of the SI model. In this respect, we  build on
the recent contribution of   \cite{lee2019transient}  who provide an analytical  framework to represent the transient dynamics of the SI epidemic dynamics on an arbitrary network. In particular, they  derive a tight approximation in closed-form of the solution to the SI epidemic dynamics over all time $t$. The latter overcomes the shortfalls of the existing linearised approximation (see \citet{canright2006spreading,mei2017dynamics,newman2010networks}) by means of a thorough mathematical transformation of the system governing the SI dynamics. \cite{lee2019transient}  have also derived vaccination policies to mitigate the risks of potential attacks or minimise the consequences of an existing epidemic spread with a limited number of available patches or vaccines over the network.

From an economic perspective, our contribution relates to the growing literature on the private provision of public goods on network. This literature mostly focuses on the relationship between the network structure and the individual provision of public goods. It generally considers a fixed network and that the public good/effort provision of an agent only affects its neighbors. In particular,  \cite{allouch2015private} shows the existence of a Nash Equilibrium in this setting under very general conditions. \cite{bramoulle2007public}  prove, in a more specific setting, that Nash Equilibria generically have a specialised structure in which some individuals contribute and others free ride.  A more recent contribution by  \cite{kinateder2017public} extends the models of private provision of public goods to a setting  with an endogenous network formation process. Yet, the network is formed in view of the benefits provided by the public good/effort offered by connections. Hence, although related, our focus differs from this strand of literature as, in our setting, the  process of link formation per se is the source of external effects, and  effects propagate throughout the network.  Another related contribution is \cite{elliott2019network}, which provides a more conceptual view on the relationship between the network structure and public goods. It focuses on the network of external effects per se and characterises efficient cooperation/bargaining institutions in this framework. Our model could be subsumed into an extended version of their model which considers multi-dimensional actions. However, their framework abstracts away from the process underlying the interactions, which is one of our key focus.

\section{The model}\label{se model}

\subsection{Notations} \label{notation}
 We consider $\mathbb{N}$ the set of natural numbers and $N\in\mathbb{N}$. The notation $\mathbb{M}^N$ (resp. $\mathbb{M}^N(\R_+)$) denotes the set of $N-$dimensional square matrices with coefficients in $\mathbb{\R}$ (resp. $\R_+$).
For a given $M\in\mathbb{M}^N$, we write $(M)_{i,j}$ or $m_{i,j}$, $1\le i,j\le N$, to refer to its element in the $i^\text{th}-$row and $j^\text{th}-$column. Moreover, for any $M\in\mathbb{M}^N$, $||M||$ denotes its Frobenius norm and for any matrix $M$ and $K$ in $\mathbb{M}^N\times \mathbb{M}^N$, we write
$M\le K\mbox{ if }m_{i,j}\le k_{i,j},\,\forall i,j=1,...,N\,.$  Additionally, the matrix $I$ (resp. $O$) stands for the $N-$dimensional square identity (resp. null) matrix.

Similarly, for a $N-$dimensional column vector $\u\in\R^N$, $u_i,\,1\le i\le N,$ refers to its element in the $i^\text{th}-$row
while $\u^\top$  denotes its transpose and $||\u||$ its Euclidean norm.
Additionally, for any $\u$ and $\v$ in $\R^N\times \R^N$, we let $\u\preceq\v\mbox{ if }u_i\le v_i,\,\forall i=1,...,N\,.$
We define similarly $\u\prec\v.$   For a function $f:\R\mapsto\R$ and a vector $\u\in\R^N$, $f(\u)$ denotes the $N-$dimensional column vector with $f(u_i), \,1\le i\le N,$ as entries. Moreover, $\mathbbm{1}$ is the $N-$dimensional column vector with one as entries.

We also consider $\text{diag}(\u)$, the $N-$dimensional square diagonal matrix with $u_i, \,1\le i\le N,$ as diagonal entries.
Additionally, for the $i^\text{th}-$vector of the canonical basis of $\R^N$, $\textbf{e}^i, \,1\le i\le N,$  and any matrix $M\in\mathbb{M}^N$, we define the product operator $$<\textbf{e}^i,M>\defeq\left(\sum_{j=1}^N\textbf{e}^i_j\times m_{j,1},...,\sum_{j=1}^N\textbf{e}^{i}_j\times m_{j,N}\right)\,,$$ a $N-$dimensional row vector.

We define $\mathbb{S}^N$ (resp. $\mathbb{S}^N(\R_+)$) as the subset of elements of $\mathbb{M}^N$ (resp. $\mathbb{M}^N(\R_+)$) that are symmetric. We observe that  $\mathbb{S}^N$ is a real vector-space of dimension ${N(N+1)}/{2}$ and we consider the basis formed by the matrices $(B^{ \{i,j \}})_{1 \leq i \leq j \leq N}$ such that $b^{\{i,j\}}_{i,j}=b^{\{i,j\}}_{j,i}=1$ and  $b^{\{i,j\}}_{k,\ell}=0$ for $\{k,\ell\} \not = \{i,j\}.$  Accordingly, given a matrix $D\in\mathbb{S}^N$, we let $d_{\{j,k\}}\defeq d_{j,k}+d_{k,j}.$ Moreover, given $U \subseteq \mathbb{S}^N$, a differentiable function $\phi:U \rightarrow  \R,$ and $\bar D \in U,$ we denote by $\dfrac{\partial \phi}{\partial d_{\{i,j\}}} (\bar D)$ the partial  derivative in the direction of $B^{\{i,j\}},$ that is
$$\dfrac{\partial \phi}{\partial d_{\{i,j\}}} (\bar D):=\dfrac{\partial \phi}{\partial d_{i,j}} (\bar D) +\dfrac{\partial \phi}{\partial d_{j,i}} (\bar D)\,, $$ where $\dfrac{\partial \phi}{\partial d_{i,j}} (\bar D)$ and $\dfrac{\partial \phi}{\partial d_{j,i}} (\bar D)$ denote the partial derivatives in the directions induced by the canonical basis of $\mathbb{M}^N.$

 Finally, for a set $\mathcal{B}$, we note $\text{Card}(\mathcal{B})$ its cardinal and $\left(\mathcal{B}\right)^\text{c}$ the complementary set.

 \subsection{Model outline}

We consider a finite set of agents, $\mathcal{N}=\{1,2,...,N\},\,N\ge 2,$ connected through a weighted and undirected network. The set of links is given by $\mathcal{E} \subseteq \{\{i,j\} \mid i,j \in \mathcal{N}\}$ and their weights by the weighted adjacency matrix $A\in \mathbb{S}^N(\R_+)$.  In particular, for all $i\in\mathcal{E}$, $a_{ii}=0$. The agents face the risk of shifting from a good/susceptible state to a bad/infected state. This transition occurs in continuous time through an epidemic process over the network. At time zero, a subset of agents idiosyncratically shifts to the infected state. Following this initial shock, infected agents contaminate their neighbours in the network with a probability that is proportional to the weight of the corresponding link. Infected agents remain so permanently and cannot revert to the susceptible state. As intimated in Section \ref{se lit}, this model is known as the SI model in the epidemiological literature (see, e.g., \cite{pastor2015epidemic}). It provides an accurate description of epidemic dynamics at short time scale and/or when no vaccine or treatment is available against the epidemic.

We consider a socio-economic setting in which strategic agents can invest in the network  in order to reduce contagion rates.  We are concerned with the characterisation of the equilibrium behaviour in this context, its relation to social efficiency, and the potential impacts of policy on these features. Such setting captures the behaviour of countries facing the global propagation of an epidemic as well as that of individuals facing its local propagation. It can also be applied to other socio-economic context such as the propagation of computer viruses (see, e.g., \cite{pastor2001epidemic}) or financial distress (see, e.g., \cite{battiston2012debtrank}).

 In order to formally define the model, we first provide a detailed description of the epidemic dynamics and its approximation (see Section \ref{epidyn}) and then introduce a representation of agents' prophylactic behaviours  (see Section \ref{behavior}).

\subsection{Epidemic dynamics}
\label{epidyn}
Formally, an exact model of the dynamics of epidemic spreading in the SI framework is given by a continuous-time Markov chain  $(X(t))_{t \geq 0}$ with state space $\mathcal{X}:=\{0,1\}^N.$ A state $X(t) \in \mathcal{X}$ is defined by the combination of states of all the $N$ agents at time $t$ and is thus described by the set of susceptible agents $\{ i \in \mathcal{N} \mid X_i(t)=0\}$ and the set of infected agents $\{ i \in \mathcal{N} \mid X_i(t)=1\}.$  The key specificity of the dynamics of the Markov chain $(X(t))_{t \geq 0}$ is  the stochastic rate of contagion for a susceptible node $i\in\mathcal{N}$, characterising its transition probability as follows
\begin{equation}  \lim_{h \rightarrow 0} \dfrac{1}{h}\mathbb{P}[X_i(t+h)=1 \mid X_i(t)=0] = \beta \sum_{j \in \mathcal{N}} a_{i,j} \mathbbm{1}_{\{X_j(t)=1\}}\,,\label{rate}\end{equation}
 where $\beta$ is a unit contagion rate, $a_{i,j},\,i,j\in\mathcal{N},$ is the contagiousness of the network link $\{i,j\}$, and $\P$ denotes the probability on the underlying probability space.
Equation  \eqref{rate} characterises completely the dynamics, as infected nodes remain so permanently. It highlights the role of the network in the contagion process and the possible heterogeneous contagiousness of different network links. In this respect, we make the following assumption about the network structure throughout the paper.
\begin{assump}\label{network}  The adjacency matrix $A\in\mathbb{S}^N(\R_+)$ is irreducible and aperiodic.
\end{assump}
The irreducibility assumption  amounts to considering that every agent faces a risk of contagion as soon as at least one agent in the network is infected. Indeed, the network is then necessarily connected and  the asymptotic behaviour of the Markov chain is trivial: there is an unstable steady state  where none of the agent is infected and a unique stable steady state where all agents are contaminated\footnote{Stability must be understood in the sense that, for any initial non-null probability distribution, the limiting distribution of the Markov chain has full support on the full contamination state.}. In the following, we shall actually consider that agents are concerned by the time at which they are likely to be infected rather than by their asymptotic infection status.
Accordingly, we are concerned with the transient behaviour of the Markov chain. However, as intimated above, the infection rate defined by the right-hand side of Equation \eqref{rate} is a random variable, making the process doubly stochastic. The random nature of the infection rate could then be offset by conditioning with respect to all the possible combinations of the neighbouring states, i.e. $X_j(t),\,j\neq i.$ Yet, such conditioning of every node would lead to a Markov chain with $2^N$ states, i.e. with a number of states increasing exponentially with the number of nodes, and thus being neither analytically nor computationally tractable.  Therefore, the conventional practice in epidemiological modelling is to consider a mean-field approximation of the infection rate. In particular, the $N$-intertwined model of \citet{van2009virus} considers the average behaviour, i.e. the expectation, over states for the infection rate. Therefrom, the latter model derives an approximate dynamics of the probability for an agent $i \in \mathcal{N}$ to be in the infected state at time $t$ given by $x_i(t):=\mathbb{P}[X_i(t)=1]$\footnote{Let $T_i\defeq\inf\{t\ge 0:\,X_i(t)=1\}$ valued in $\R_+$.
We observe that for $t\ge 0$, $x_i(t)=\p[T_i\le t]$.}. More precisely, using the fact that $\mathbb{P}[X_i(t)=1]+\mathbb{P}[X_i(t)=0]=1$ and the total law of probability,  one gets from Equation \eqref{rate} that for all  $i \in \mathcal{N},$
\begin{equation} \dfrac{\partial x_i(t)}{\partial t} = (1-x_i(t)) \beta \sum_{j=1}^N a_{i,j} x_j(t)\,.  \label{vm} \end{equation}
Equation \eqref{vm} thus provides a deterministic approximation of the dynamics of the contagion probability that takes  into account the full network structure. It nevertheless disregards the positive correlation between the infection status of neighbouring nodes. This implies that Equation \eqref{vm} over-estimates the probability of contagion (see \cite{van2009virus}).

\begin{rem} Alternative mean-field approximations used in the literature are generally much coarser that the $N$-intertwined model considered here. Two common approaches are (i) to average over agents and focus on the (approximate) dynamics of the average probability of contagion  or (ii) to average over agents with equal degree  and focus on the (approximate) dynamics of the average probability of contagion for an agent of a given degree (see \cite{pastor2015epidemic} for an extensive review).
\label{others}\end{rem}

The non-linear Equation \eqref{vm} does not have an analytical solution. A common approach in the literature, used in particular to analyse the outbreak of an epidemic, is to assume $x_i(t)$ small enough to discard the factor $(1-x_i(t))$  and thus focus on the following linear equation
\begin{equation} \dfrac{\partial x_i(t)}{\partial t} = \beta \sum_{j=1}^N a_{i,j} x_j(t) \,. \label{linear} \end{equation}
However, this approximation grows exponentially towards $+ \infty,$ whereas it is assumed to approximate a probability. In a recent contribution,  \citet{lee2019transient} provide a much better approximation of the solution of Equation \eqref{vm}. More precisely, they define for all $i \in \mathcal{N}$ and $t \in \R_+,$ $y_i(t):=- \log(1-x_i(t)),$ and observe that $\bar{\x}:=[\bar x_1,...,\bar x_N]^\top$ is a solution  of the system defined by Equation \eqref{vm} with initial condition $\x(0):=[x_1(0),...,x_N(0)]^\top=:\x_0$, with at least one non-null element to avoid triviality, if and only if $\bar{\y}:=[\bar y_1,...,\bar y_N]^\top$ is a solution of the system of equations defined for all $i\in \mathcal{N}$ by
\begin{equation} \dfrac{\partial y_i(t)}{\partial t} = \beta \sum_{j \in \mathcal{N}} a_{i,j}(1-\exp(-y_j(t)))\,,  \label{eqy1} \end{equation}
with the corresponding initial condition.
They then show that a tight upper bound  to the solution of the system defined by Equation \eqref{eqy1} when $\x(0)=\x_0\prec 1$ is provided by
\begin{equation}
\check{\y}(t)\defeq-\ln(1-\x_0)+\left[\exp{(\beta tA\text{diag}(1-\x_0))}-I\right]\text{diag}(1-\x_0)^{-1}\x_0\,,
\label{approx} \end{equation}
 and accordingly that $\check{\x}(t)\defeq 1- \exp{(-\check{\y}(t))}$ is a tight upper bound to the solution $\bar{\x}$ of the system defined by Equation \eqref{vm} with initial condition $\x(0)=\x_0$
 in the sense that  one has (see \citet[Theorem 5.1 and Corollary 5.2]{lee2019transient})
 \begin{itemize*}
 \item  $\lim_{t \rightarrow + \infty} ||\check{\x}(t)-\x(t)|| = 0,$
 \item  for any $t\ge 0$, $\bar \x(t)\preceq \check{\x}(t)\preceq \tilde{\x}(t)$ where $\tilde{\x}:=[\tilde x_1,...,\tilde x_N]^\top$ is the solution of the system defined by Equation \eqref{linear} with initial condition $\x(0)=\x_0.$
 \end{itemize*}
 Hence, $ \check{\x}$ provides an approximation of the probability of contagion that is asymptotically exact and more accurate than the standard linear approximation, even at short time scale.

 \subsection{Prophylactic behaviour}
 \label{behavior}

From now on, we shall consider that agents base their assessment of the dynamics of contagion on the approximated  contagion probabilities  $ \check{\x}$ associated to a given and fixed initial condition $\x(0)=\x_0\prec 1$, having at least one non-null element.  In this sense, they make decisions on the basis of approximate information. This approach provides a consistent representation of the decision-making situation of actual agents which ought to base their decisions on similar approximations.

 In this respect, we recall that in our SI setting, all agents eventually become infected. Thus, agents cannot base their decisions on their asymptotic infection status. Rather, they shall aim at delaying the growth rate of the epidemic. This is notably the strategy pursued by most countries during the recent COVID-19 pandemic.  More precisely, we consider that agents consider a target date $\bar{t},$ which can be interpreted as the planning horizon or the expected date of availability of a treatment, and aim at minimising the probability of contagion up to that date. We further assume, for sake of analytical tractability, that they have a logarithmic utility of the form
 \begin{equation*} u_i(\check{x}_i(\bar{t})) := \delta_i\log(1-\check{x}_i(\bar{t})),\,i\in\mathcal{N}\,, \end{equation*}
where $\check{x}_i(\bar{t})$ is the approximate contagion probability given by Equation  \eqref{approx} and $\delta_i\ge 0$ is a subjective measure of the value of avoided contagion, or equivalently of the  cost of contagion, for the agent $i$. One should note that the utility is non-positive and equal to a benchmark of zero if and only if there is no  risk of contagion.  In our setting, $\x_0$, $\beta$, and $\bar t$ being fixed, Equation \eqref{approx} implies that the contagion probability is completely determined by the adjacency matrix $A$. The utility of agent $i\in\mathcal{N}$ can thus be expressed directly as
\begin{equation} v_i(A):=-\delta_i<\textbf{e}^i,\exp{(\beta \bar tA\text{diag}(1-\x_0))}\text{diag}(1-\x_0)^{-1}\x_0> \,, \label{netutil} \end{equation}
where the constant term $\ln(1-\x_0)+\text{diag}(1-\x_0)^{-1}\x_0$ has been discarded to simplify the notations.

Equation \eqref{netutil} highlights that, for a given admissible initial probability of contagion $\x_0$, the only lever that agents can use to reduce their contagion probability is the decrease of the contagiousness of the network, i.e. the decrease of the value of the coefficients of the adjacency matrix $A.$ This is exactly the strategy put in place during the COVID-19 pandemic, at the local scale through social distancing measures, and at the global scale through travel restrictions and border shutdowns (see \cite{colizza2006role} for an analysis of the role of the global transport network in epidemic propagation). Formally, we consider a strategic game in which each agent can invest in the reduction of contagiousness of network links.
We distinguish two alternative settings to account for potential constraints on  agents' actions:
\begin{itemize*}
\item In the global game, we assume that each agent can invest in the reduction of contagiousness of every network link. Therefore, the set of admissible strategy profiles is given by
$\mathcal{S}(A)\defeq\{(D^i)_{i\in\mathcal{N}} \in (\mathbb{S}^N(\R_+))^{N}:\,A-\sum_{i\in\mathcal{N}}D^i\ge 0\}.$
\item In the local game, we assume that each agent can only invest in the links through which it is connected.  Therefore, the set of admissible strategy profiles is given by
$\mathcal{K}(A)\defeq\{(D^i)_{i\in\mathcal{N}}\in\mathcal{S}(A):\,\forall\,i\in\mathcal{N},\,\forall\,k,j\in\mathcal{N}, \,k,j\neq i \Rightarrow d_{k,j}^i=0\}.$
\end{itemize*}
Local games correspond to a setting where agents are individuals that limit their social interactions through individual and costly measures. On the other hand, global games apply to a more involved setting where agents are organisations (regions, countries) that have the ability to subsidise the investment of other agents in the reduction of contagiousness, either directly or indirectly.

\begin{rem}\label{re setprop}
Both $\mathcal{S}(A)$ and $\mathcal{K}(A)$ are non-empty, convex and compact sets.
\end{rem}

The payoff function is defined in a similar fashion in both settings:
\begin{itemize*}
\item First, a strategy profile $(D^i)_{i \in \mathcal{N}}$ turns the adjacency matrix into  $A -\sum_{i \in \mathcal{N}} D^i$ and thus yields to agent $i \in \mathcal{N}$ a utility
\begin{small}
\begin{align*} \hspace{-1cm} U_i(D^i,D^{-i})&:= v_i(A -\sum_{i \in \mathcal{N}} D^i)= -\delta_i<\textbf{e}^i,\exp{\left(\beta \bar t(A -\sum_{i \in \mathcal{N}} D^i)\diago(1-\x_0)\right)}\diago(1-\x_0)^{-1}\x_0> \,,\end{align*}
\end{small}
where $D^{-i}:=(D^j)_{j \in \mathcal{N},\,j\neq i}$ is the strategy profile of all agents but $i$.
\item Second, we consider that agents face a linear cost for their investment in the reduction of contagion. More precisely, one has for all $i\in\mathcal{N},$
\begin{small}
\begin{equation*}C_i(D^i):= \rho \mathbbm{1}^\top D^i\mathbbm{1}=\rho\sum_{j,k\in\mathcal{N}} d^i_{j,k}\,, \end{equation*}
\end{small}
where $\rho>0$ is the cost parameter.
\item Overall, the payoff of agent $i \in \mathcal{N}$ given a strategy profile $(D^i)_{i \in \mathcal{N}}$ is given by
\begin{small}
\begin{align*} \hspace{-2em}\Pi_i(D^i,D^{-i})& := U_i(D^i,D^{-i})-C_i(D^i) \\ &= -\delta_i<\textbf{e}^i,\exp{\left(\beta \bar t(A -\sum_{i \in \mathcal{N}} D^i)\diago(1-\x_0)\right)}\diago(1-\x_0)^{-1}\x_0> -\rho\mathbbm{1}^\top D^i\mathbbm{1}\,.
\end{align*}
\end{small}
\end{itemize*}
A few remarks are in order about the characteristics of the game.  First,  agents' strategy sets are constrained by the choices of  other players. Namely, given a strategy profile for the other players $D^{-i} \in (\mathbb{S}^N(\R_+))^{N-1}$,  the set of admissible strategies for player $i$  is $\mathcal{S}_i(A,D^{-i}):=\{D^i \in \mathbb{S}^N(\R_+) \mid (D^i,D^{-i}) \in \mathcal{S}(A) \}$  (resp. $\mathcal{K}_i(A,D^{-i}):=\{D^i \in \mathbb{S}^N(\R_+) \mid (D^i,D^{-i}) \in \mathcal{K}(A) \}$) in the global (resp. local) game.  Although, it is not the most standard, this setting is comprehensively analysed in the literature (see, e.g., \cite{rosen1964existence}).  Second, linear cost is  a natural assumption in our framework. Indeed, the marginal cost paid to decrease the contagiousness of a link should not depend on the identity of the player investing.
Third, the payoff function is always non-positive as it is the combination of both a  utility and a cost that are always non-positive. Finally, for large $\bar t$ and homogeneous initial contagion probabilities, the utility can be approximated through the eigenvector centrality of the contagion network, as detailed in the following lemma.

\begin{dfn}[$(\alpha)$-Homogeneous Game]
 A game is $(\alpha)$-homogeneous if it involves an homogeneous initial probability of contagion, i.e. such that for all $i\in\mathcal{N}$, $x_{0_i}= \alpha$ for some $\alpha\in(0,1)$.
\end{dfn}

\begin{lem}\label{le eigenvec} Consider an $(\alpha)$-Homogeneous Game, and let $D \in \mathcal{S}(A)$ be such that $A-\sum_{i \in \mathcal{N}} D^i$ is irreducible and aperiodic. Let then $\mu_1$ denote the Perron-Frobenius eigenvalue of $(A-\sum_{i \in \mathcal{N}} D^i),$  $|\mu_1-\mu_2 |$ the spectral gap and $\v$ the normalised eigenvector associated to $\mu_1,$ corresponding to the eigenvector centrality of the network. One has
\small
\begin{equation*}
\exp{\left(\beta \bar t(1-\alpha)(A -\sum_{i \in \mathcal{N}} D^i)\right)}= \left(1+ \mathcal{O}(\exp( -\beta \bar t(1-\alpha)|\mu_1-\mu_2 | )) \right)   \exp (\beta \bar t(1-\alpha)\mu_1 )  \v \v^\top\,. \end{equation*}
\end{lem}
\normalsize
\begin{rem}
As $A$ is irreducible and aperiodic, the condition $\sum_{i \in \mathcal{N}} d_{j,k}^i < a_{j,k}$ for all $j,k\in\mathcal{N}$ such that $a_{j,k}>0,$ is sufficient for getting the irreducibility and aperiodicity of $A-\sum_{i \in \mathcal{N}} D^i$ .
\end{rem}

We now state some properties on the marginal utility and the payoff function (recall the notations
for the partial derivatives of a symmetric matrix in  Section \ref{notation}).

\begin{lem}\label{le marginal}
For every $i \in \mathcal{N},$ for any strategy profile $(D^i, D^{-i}) \in \mathcal{S}(A),$ and for all $k,\ell \in \mathcal{N},$
\begin{align}\dfrac{\partial U_i(\cdot, D^{-i})} {\partial d^i_{\{k,\ell\}}}(D^i)&=   \dfrac{\partial U_i(\cdot, D^{-i})} {\partial d^i_{k,\ell}}(D^i) +  \dfrac{\partial U_i(\cdot, D^{-i})} {\partial d^i_{\ell,k}}(D^i)\nonumber \\
&=\delta_i \beta \bar t \left(\exp(\beta \bar t(A-\sum_{j\in\mathcal{N},\,j \not = i}  D^j- D^i)\diago(1-\x_0))\right)_{i,k} x_{0_\ell}\nonumber\\
 &\quad+  \delta_i \beta \bar t \left(\exp(\beta \bar t(A-\sum_{j\in\mathcal{N},\,j \not = i}  D^j- D^i)\diago(1-\x_0))\right)_{i,\ell} x_{0_k}\,,\label{eq der}
\end{align}
and the marginal utility is non-negative. Moreover, the map $\Pi_i(\cdot, D^{-i})$  is concave on $\mathcal{S}_i(A, D^{-i}).$
\end{lem}
From the previous lemma derives the following result relating the marginal utility of every agent to the utility itself, when the initial probability of infection is constant among agents.

\begin{lem}\label{le equivutility}
Consider an $(\alpha)$-Homogeneous Game, and let $\mathcal{M}\subseteq\mathcal{N}$. For every $i \in \mathcal{N},$ and for any strategy profile $(D^i, D^{-i}) \in \mathcal{S}(A),$
\begin{align}\label{eq derveq}
\sum_{\substack{k,\ell\in\mathcal{N}\\k\,\text{or}\,\ell\in\mathcal{M}}}\dfrac{\partial U_i(\cdot, D^{-i})} {\partial d^i_{\{k,\ell\}}}(D^i)&=-2{\rm Card}(\mathcal{M})\delta_i\beta\bar t \alpha \sum_{k\in\mathcal{N}}\left(\exp(\beta \bar t(1-\alpha)(A-\sum_{i\in\mathcal{N}}  D^i))\right)_{i,k}\nonumber\\&=-2{\rm Card}(\mathcal{M})\beta\bar t(1-\alpha) U_i(\cdot, D^{-i})(D^i)\,.
\end{align}
\end{lem}

\subsection{Nash Equilibrium and Social Optimum}

In the following, unless otherwise specified, we consider as implicitly given the utility weights ${\bdelta}:=[\delta_1,...,\delta_N]^\top,$ the time-horizon $\bar t,$ the unit contagion rate $\beta$, the initial contagion matrix $A$, the initial contagion probabilities $\x_0,$ and the investment cost $\rho$.  We then define the ``local game" $\mathcal{L}(\bdelta, A,\beta,\bar{t},\x_0,\rho)$  as the game  with  strategy profiles in $\mathcal{K}(A)$ and  payoff function $\Pi$ and  the ``global game" $\mathcal{G}(\bdelta,A,\beta,\bar{t},\x_0,\rho)$  as the one with strategy profiles in $\mathcal{S}(A)$ and  payoff function $\Pi.$  In this setting, a Nash Equilibrium is defined as follows.

\begin{dfn}[Nash Equilibrium]\label{def nash}	 \
\begin{itemize*}
\item An admissible set of strategies $\check D\defeq(\check D^i)_{i\in\mathcal{N}}\in  \mathcal{S}(A)$ is a Nash Equilibrium for the global  game  if
\begin{equation*}
\forall i \in \mathcal{N}, \, \forall D^i \in \mathcal{S}_i(A,\check D^{-i}),\,  \Pi_i(\check D^i,\check D^{-i}) \geq \Pi_i( D^i,\check D^{-i})\,. \end{equation*}
\item An admissible set of strategies $\bar D\defeq(\bar D^i)_{i\in\mathcal{N}}\in  \mathcal{K}(A)$ is a Nash Equilibrium for the local  game  if
\begin{equation*}
\forall i \in \mathcal{N}, \, \forall D^i \in \mathcal{K}_i(A,\bar D^{-i}),\,  \Pi_i(\bar D^i,\bar D^{-i}) \geq \Pi_i( D^i,\bar D^{-i}) \,.\end{equation*}
\end{itemize*}
\end{dfn}

The existence of a Nash Equilibrium follows from standard arguments.

\begin{theo}\label{th NEEx}
There exists a Nash Equilibrium in both the local and global games.
\end{theo}

\begin{rem}
In our setting, equilibrium is in general not unique as there might be indeterminacy on the identity of the players/neighbours which ought to invest in reducing the contagion of a link (see the discussions in Section \ref{se eq} below).
\end{rem}

The key concern, in the remaining  of this paper, is the study of the efficiency of  Nash Equilibrium. As commonly done in $N$-agent games, and in particular in network games, we define as Social Optimum, the outcome that maximises the equally-weighted sum of individual utilities.

\begin{dfn}[Social Optimum]\label{def sosopt}
An admissible set of strategies $\hat{D}\defeq(\hat D^i)_{i\in\mathcal{N}}\in{\mathcal{S}}(A)$ is a Social Optimum if
\begin{align*}
\hat D={\rm argmax}_{(D^i)_{i\in\mathcal{N}}\in\mathcal{S}(A)}\sum_{i\in\mathcal{N}}\Pi_i \left(D^{-i},D^i\right)\,.
\end{align*}
\end{dfn}
Note that $\sum_{i\in\mathcal{N}}\Pi_i(D^i,D^{-i})$ only depends on the value of $\sum_{i\in\mathcal{N}} D^i.$ First, this implies that
 the notion of Social Optimum is the same in the local and global game.  Indeed,  it is straightforward to check that for every $(D^i)_{i\in\mathcal{N}}\in\mathcal{S}(A),$ there exists $(\tilde{D}^i)_{i\in\mathcal{N}}\in\mathcal{K}(A)$ such that  $\sum_{i\in\mathcal{N}} \tilde{D}^i= \sum_{i\in\mathcal{N}} D^i.$ Second, given a matrix $D\in\mathbb{S}^N(\R_+)$ such that $ D \leq A, $ we shall let
 \begin{equation*} \hat \Pi(D):= \sum_{i \in \mathcal{N}} \hat v_i(D)- \rho \sum_{j,k\in\mathcal{N}} d_{j,k}\,,  \end{equation*}
where $\hat v_i:D\mapsto v_i(A-D)$, and, with a slight abuse of notation, state that $\hat D$ is a Social Optimum if it is such that $\hat \Pi(D)$ is maximal over $\mathcal{D}(A):=\{D \in \mathbb{S}^N(\R_+):\,  D \leq A\}$. The existence of a Social Optimum directly follows from the continuity of $\hat \Pi$ and the compactness of $\mathcal{D}(A)$ .

\begin{theo}\label{th exisoc}
There exists a Social Optimum in both the local and global games.
\end{theo}

We end this section with two lemmas that are the counterparts of Lemma \ref{le marginal}-\ref{le equivutility} above for the Social Optimum $D\in \mathcal{D}(A)$ and whose proofs are a straightforward adaptation of the proofs of the latter lemmas.
\begin{lem}\label{le marginalbis}
For every $i \in \mathcal{N},$ for any Social Optimum $D\in \mathcal{D}(A),$ and for all $k,\ell \in \mathcal{N},$
\begin{align*}\dfrac{\partial \hat v_i(D)} {\partial d_{\{k,\ell\}}}&= \delta_i \beta \bar t \left(\exp(\beta \bar t(A-D)\diago(1-\x_0))\right)_{i,k} x_{0_\ell}\nonumber\\
 &\quad+  \delta_i \beta \bar t \left(\exp(\beta \bar t(A-D)\diago(1-\x_0))\right)_{i,\ell} x_{0_k}\ge 0\,.
\end{align*}
Moreover, the map $\hat v_i(\cdot)$  is concave on $\mathcal{D}(A).$
\end{lem}

\begin{lem}\label{le equivutilitybis}
Consider an $(\alpha)$-Homogeneous Game, and let $\mathcal{M}\subseteq\mathcal{N}$. For every $i \in \mathcal{N},$ and for any Social Optimum $D\in \mathcal{D}(A),$
\begin{align}\label{eq derveq2}
&\sum_{\substack{k,\ell\in\mathcal{N}\\k\,\text{or}\,\ell\in\mathcal{M}}}\dfrac{\partial \hat v_i(D)} {\partial d_{\{k,\ell\}}}=-2{\rm Card}(\mathcal{M})\beta\bar t(1-\alpha)\hat v_i(D)\,.
\end{align}
\end{lem}
\section{Individual behaviour and policy response}\label{se eqsoc}
In this section, we provide a characterisation of equilibrium behaviours as a function of the network structure. We then measure, through the PoA, potential inefficiencies induced by individual behaviours, giving insight on how to restore optimality in the local game.

\subsection{Characterisation of equilibrium behaviours}\label{se eq}
Let $i\in\mathcal{N}$. Individual prophylactic efforts are characterised by the investment in contagion reduction $D^i \in \mathbb{S}^N(\R_+)$ that the individual performs in the links it has access to ($\mathcal{S}(A)$  or  $\mathcal{K}(A)$). As emphasised above, the network of contagion is assumed undirected and thus  investments in the link $\{k,\ell\}\in\mathcal{E}$ induce an equal reduction of the contagion weights $a_{k,\ell}$ and $a_{\ell,k}.$  More precisely, the marginal impact of the investment made by player $i$ is characterised by Equation \eqref{eq der} above. This equation highlights the bilateral impact of agents' investments on contagion from  $\ell$ to $k$ on the one hand and from $k$ to $\ell$ on the other hand.  The impact on agent $i$ of reduced contagiousness through $d^i_{k,\ell}$  (from $\ell$ to $k$) depends on (i) the utility weight $\delta_i$, (ii) the initial contagiousness of node $\ell,$ $x_{0_\ell},$ and (iii) the connectivity between $k$ and $i$ measured, for $D:=(D^i)_{i\in\mathcal{N}}$, through $C_{i,k}(A,D,\beta,\bar t,\x_0):=\beta \bar t  \left(\exp(\beta \bar t(A-\sum_{j\in\mathcal{N}}  D^j)\diago(1-\x_0))\right)_{i,k},$ or alternatively
\begin{equation*}  C_{i,k}(A,D,\beta,\bar t,\x_0) = \frac{1}{\fact{n}}\sum_{n=0}^\infty(\beta\bar t)^{n+1}\left( (A-\sum_{j \in \mathcal{N}}  D^j) \diago(1-\x_0)\right)^n_{i,k}\,.\end{equation*}
Hence, the connectivity from $k$ to $i,$ $C_{i,k}(A,D,\beta,\bar t,\x_0),$  depends on the initial structure of the contagion network $A,$ the strategic investments in the reduction of contagiousness $D$, the unit contagion rate $\beta$, the time-horizon $\bar t,$ and the initial contagion probabilities $\x_0$. It corresponds to the discounted sum of paths from $k$ to $i$ where each path is assigned a weight corresponding to the probability that it  propagates a genuine contagion sequence. The weight of each link across the path is thus obtained  as the product of (i) the probability that the link propagates the contagion, which is measured through the corresponding coefficient $(A-\sum_{j\in\mathcal{N}}  D^j),$ and (ii)  the probability that the end node is not initially infected, which is measured  through the corresponding coefficient of  $\diago(1-\x_0).$   Overall, the marginal  impact of agents' actions on contagiousness depends on the characteristics of the disease, measured through the initial contagion probability, and on the structure of the contagion network modified by the agents' investments.
Equation \eqref{eq der} offers a differential characterisation of Nash Equilibria in both the local and global games, as reported in the following two propositions.

\begin{prop}\label{prop NELocal}
A strategy profile $\bar D\in\mathcal{K}(A)$ is a Nash Equilibrium of the local game $\mathcal{L}(\bdelta, A,\beta,\bar{t},\x_0,\rho)$  if and only if for all $\{k,\ell\} \in\mathcal{E},$ the following two conditions hold:
\begin{enumerate}[wide =0pt, label=(\arabic*)]
\setlength{\itemsep}{0pt}
\setlength{\parskip}{0pt}
\setlength{\parsep}{0pt}
\item One of the following alternative holds:
\small
\begin{enumerate}[wide =0pt, label=(\alph*)]
\setlength{\itemsep}{0pt}
\setlength{\parskip}{0pt}
\setlength{\parsep}{0pt}
\item $\rho<\max_{i \in \{k,\ell\}} \delta_i\left(  C_{i,k}(A,\bar D,\beta,\bar t,\x_0) x_{0_\ell} +C_{i,\ell}(A,\bar D,\beta,\bar t,\x_0)  x_{0_k}  \right)$ and $\bar d^k_{\{k,\ell\}}+\bar d^\ell_{\{k,\ell\}}=a_{\{k,\ell\}},$
    \item $\rho>\max_{i \in \{k,\ell\}} \delta_i\left(  C_{i,k}(A,\bar D,\beta,\bar t,\x_0) x_{0_\ell} +C_{i,\ell}(A,\bar D,\beta,\bar t,\x_0)  x_{0_k}  \right)$ and $\bar d^k_{\{k,\ell\}}=\bar d^\ell_{\{k,\ell\}}=0,$
\item $\rho=\max_{i \in \{k,\ell\}} \delta_i\left(  C_{i,k}(A,\bar D,\beta,\bar t,\x_0) x_{0_\ell}+C_{i,\ell}(A,\bar D,\beta,\bar t,\x_0) x_{0_k}  \right)$ and $\bar d^k_{\{k,\ell\}}+\bar d^\ell_{\{k,\ell\}}\in [0,a_{\{k,\ell\}}]$.
\end{enumerate}
\normalsize
\item For any $i\in\mathcal{N}$, one has  $\bar d^i_{\{k,\ell\}}>0$ only if $$\delta_i\left(  C_{i,k}(A,\bar D,\beta,\bar t,\x_0) x_{0_\ell} +C_{i,\ell}(A,\bar D,\beta,\bar t,\x_0) x_{0_k}  \right) \geq \rho\,.$$
\end{enumerate}
\end{prop}

\begin{prop}\label{prop NEGlobal}
A strategy profile $\check D\in\mathcal{S}(A)$ is a Nash Equilibrium of the global game $\mathcal{G}(\bdelta, A,\beta,\bar{t},\x_0,\rho)$  if and only if for all $\{k,\ell\} \in\mathcal{E},$ the following two conditions hold:
 \begin{enumerate}[wide =0pt, label=(\arabic*)]
 \setlength{\itemsep}{0pt}
\setlength{\parskip}{0pt}
\setlength{\parsep}{0pt}
\item One of the following alternative holds:
\small
\begin{enumerate}[wide =0pt, label=(\alph*)]
\setlength{\itemsep}{0pt}
\setlength{\parskip}{0pt}
\setlength{\parsep}{0pt}
\item $\rho<\max_{i \in \mathcal{N}} \delta_i\left(  C_{i,k}(A,\check D,\beta,\bar t,\x_0) x_{0_\ell} +C_{i,\ell}(A,\check D,\beta,\bar t,\x_0)  x_{0_k} \right)$ and $\sum_{i\in\mathcal{N}}\check d^i_{\{k,\ell\}}=a_{\{k,\ell\}},$
    \item $\rho>\max_{i \in \mathcal{N}} \delta_i\left(  C_{i,k}(A,\check D,\beta,\bar t,\x_0) x_{0_\ell} +C_{i,\ell}(A,\check D,\beta,\bar t,\x_0)  x_{0_k}  \right)$  and $\sum_{i\in\mathcal{N}}\check d^i_{\{k,\ell\}}=0,$
\item  $\rho= \max_{i \in \mathcal{N}} \delta_i\left(  C_{i,k}(A,\check D,\beta,\bar t,\x_0) x_{0_\ell} +C_{i,\ell}(A,\check D,\beta,\bar t,\x_0) x_{0_k}  \right)$ and $ \sum_{i\in\mathcal{N}}\check d^i_{\{k,\ell\}} \in [0, a_{\{k,\ell\}}].$
\end{enumerate}
\normalsize
\item For any $i\in\mathcal{N}$, one has  $\check d^i_{\{k,\ell\}}>0$  only if $$\delta_i\left(  C_{i,k}(A,\check D,\beta,\bar t,\x_0) x_{0_\ell} +C_{i,\ell}(A,D,\beta,\bar t,\x_0)  x_{0_k} \right) \geq \rho\,.$$
\end{enumerate}
\end{prop}

\indent The difference between Proposition \ref{prop NELocal} and  \ref{prop NEGlobal} stems from the fact that different sets of agents can invest in a given link: the agents at the nodes corresponding to that link in the local case and all  agents in the global case. Otherwise, their interpretation is similar. If the investment cost is large with respect to the marginal utility of  players at equilibrium, there is no investment in the link. If the investment cost is smaller than the marginal utility of at least one player at equilibrium, there is full investment in the link, i.e. it is completely suppressed.  Finally, there is  the ``interior" case in which  only the agents with the largest marginal utility invest in the link.
They do so up to the point where their individual marginal utility equals the investment cost. The following definition summarises the different characterisations of equilibria.
\begin{dfn}[Equilibrium characterisation]\
\begin{itemize*}
\item A local (resp. global) Full Investment Equilibrium is an equilibrium that satisfies, for all  $\{k,\ell\} \in\mathcal{E},$ case (a) of Proposition \ref{prop NELocal} (resp. \ref{prop NEGlobal}).
\item A local (resp. global) No Investment Equilibrium is an equilibrium that satisfies, for all  $\{k,\ell\} \in\mathcal{E},$ case (b) of Proposition \ref{prop NELocal} (resp. \ref{prop NEGlobal}).
\item A local (resp. global) Interior Equilibrium is an equilibrium that satisfies, for all  $\{k,\ell\} \in\mathcal{E},$ case (c) of Proposition \ref{prop NELocal} (resp. \ref{prop NEGlobal}).
\item A local (resp. global) Homogeneous Interior Equilibrium is a special case of local (resp. global) Interior Equilibrium where, for each  $\{k,\ell\} \in\mathcal{E}$, the marginal utilities of agents $k,\ell$ (resp. all agents) are equal.
\end{itemize*}
\end{dfn}

We observe that in the case of a Full Investment Equilibrium or an Interior Equilibrium, there can be an indeterminacy on the identities of the agents that invest. Namely, let $E^{\mathcal{G}}_{\{k,\ell\}}( D):=\{i \in \mathcal{N} \mid \delta_i\left(  C_{i,k}(A, D,\beta,\bar t,\x_0) x_{0_\ell} +C_{i,\ell}(A, D,\beta,\bar t,\x_0)  x_{0_k}  \right) \geq \rho\},$  be the set of  players susceptible to invest in the link $\{k,\ell\} \in \mathcal{E}$ at an equilibrium $D$ of the global game and $E^{\mathcal{L}}_{\{k,\ell\}}( D):=\{i \in \{k,\ell\}\ \mid \delta_i\left(  C_{i,k}(A, D,\beta,\bar t,\x_0) x_{0_\ell} +C_{i,\ell}(A, D,\beta,\bar t,\x_0)  x_{0_k}  \right) \geq \rho\},$  be the set of  players susceptible to invest in the link $\{k,\ell\} \in \mathcal{E}$ at an equilibrium $D$ of the local game. Proposition \ref{prop indeterminacy} below, resulting from Proposition \ref{prop NELocal}-\ref{prop NEGlobal}, highlights a form of  substitutability of investments that arises at equilibrium.

\begin{prop}\label{prop indeterminacy}
Let  $D$ be an equilibrium of the global game $\mathcal{G}(\bdelta, A,\beta,\bar{t},\x_0,\rho)$ (resp. local game $\mathcal{L}(\bdelta, A,\beta,\bar{t},\x_0,\rho)$).  Assume that $\tilde D\in \mathcal{S}(A)$ (resp. $\tilde D\in\mathcal{K}(A)$) is such that for all $\{k,\ell\} \in \mathcal{E},$ one has:
 \begin{enumerate}[label=(\arabic*)]
   \setlength{\itemsep}{0pt}
\setlength{\parskip}{0pt}
\setlength{\parsep}{0pt}
 \item $\sum_{i \in \mathcal{N}} \tilde d^i_{\{k,\ell\}}=\sum_{i \in \mathcal{N}} d^i_{\{k,\ell\}}$,
  \item For any $i \in \mathcal{N},$ $\tilde d^i_{\{k,\ell\}}>0$  only if $i \in  E^{\mathcal{G}}_{\{k,\ell\}}$ (resp. $i\in E^{\mathcal{L}}_{\{k,\ell\}}$).
 \end{enumerate}
 Then $\tilde D$ is an equilibrium of the global (resp. local) game.
\end{prop}
Hence, each player that has a large enough marginal utility is willing to invest in a link up to the equilibrium level independently of the actions of other players. This  leads to indeterminacy on the allocation of investments (and thus of the related costs) among players that have a large enough marginal utility.

\begin{rem}\label{re nonullinv}
Consider the game $\mathcal{G}(\bdelta, A,\beta,\bar{t},\x_0,\rho)$ (resp. $\mathcal{L}(\bdelta, A,\beta,\bar{t},\x_0,\rho))$ and its equilibrium $D$. We observe that, whenever for some $i,j,k,\ell\in\mathcal{N}$,  $\delta_i(C_{i,k}(\cdot) x_{0_\ell} +C_{i,\ell}(\cdot)  x_{0_k} )(A, D,\beta,\bar t,\x_0) \ge\delta_j(  C_{j,k}(\cdot)x_{0_\ell} +C_{j,\ell}(\cdot)  x_{0_k} )(A, D,\beta,\bar t,\x_0)$, then $j \in  E^{\mathcal{G}}_{\{k,\ell\}}$ (resp. $j \in  E^{\mathcal{L}}_{\{k,\ell\}}$) implies $i \in  E^{\mathcal{G}}_{\{k,\ell\}}$ (resp. $i \in  E^{\mathcal{L}}_{\{k,\ell\}}$ provided that $i=k$ or $\ell$).
\end{rem}

Moreover, Proposition \ref{prop NELocal}-\ref{prop indeterminacy} imply that  Full Investment Equilibria and Interior Equilibria have a notable property: they induce equilibria in each network that is more strongly connected than the equilibrium network (i.e. with a weight on each link higher than the one of the corresponding link in the equilibrium network). This property is formally stated in the following proposition.

 \begin{prop}
 \label{prop weakint} Let $D$ be a Full Investment Equilibrium or an Interior Equilibrium of the global game $\mathcal{G}(\bdelta, A,\beta,\bar{t},\x_0,\rho)$ (resp. local game $\mathcal{L}(\bdelta, A,\beta,\bar{t},\x_0,\rho)$). Then for all $\tilde A\ge A-\sum_{i\in\mathcal{N}}D^i$,  any strategy profile $\tilde D\in\mathcal{S}(\tilde A)$ (resp. $\tilde D\in\mathcal{K}(\tilde A)$) such that $\sum_{i\in\mathcal{N}}\tilde D^{i}:= \sum_{i\in\mathcal{N}} D^i+ \tilde A-A$ is a Full Investment Equilibrium or an Interior Equilibrium  of $\mathcal{G}(\bdelta, \tilde A,\beta,\bar{t},\x_0,\rho)$ (resp. $\mathcal{L}(\bdelta, \tilde A,\beta,\bar{t},\x_0,\rho)$). Finally, both equilibria induce the same equilibrium network.
 \end{prop}
 We end this section discussing a salient example of existence of global equilibria in local strategies, that of a symmetric game and an almost fully-connected network.
\begin{dfn}[$(\alpha,\delta)$-Symmetric Game \& Complete/$(a)$-Complete Network]\
\begin{itemize*}
\item A game is $(\alpha,\delta)$-symmetric if for all $i\in\mathcal{N},$ $\delta_i=\delta$ for some $\delta>0$ and $x_{0_i}= \alpha$ for some $\alpha\in(0,1)$.
\item A network is complete if  for all $k,\ell\in\mathcal{N},\,k\neq \ell,$ $a_{k,\ell}>0$. In particular, the network is $(a)$-complete if  for all $k,\ell\in\mathcal{N},\,k\neq \ell,$ $a_{k,\ell}=a $ for some $a>0.$
\end{itemize*}
\end{dfn}

\begin{ex}
\label{ex completenet}
Consider an $(\alpha,\delta)$-Symmetric Game and an $(a)$-Complete Network.  The game is then symmetric and, using the non-emptiness, convexity and compactness properties  of the strategy space as well as the continuity and concavity properties of the payoff, there exists a symmetric equilibrium in both the local and global games (see \citet[Theorem 3]{cheng2004notes}).
We shall show that both equilibria coincide and that, for $\rho$ in an appropriate range, they are interior.
Indeed, let $\check D \in \mathcal{S}(A),$ be a symmetric equilibrium of the global game. According to Equation \eqref{eq der}, one has for all $i,k,\ell\in\mathcal{N},$
\begin{align}\label{eq derb}\dfrac{\partial U_i(\cdot,\check D^{-i})} {\partial d^i_{\{k,\ell\}}}(\check D^i)= \delta\beta\bar t\alpha\left(\exp(H)_{i,k}+\exp(H)_{i,\ell}\right)\,,\end{align}
where $H$ is of the form
\begin{align}\label{eq matH}
H\defeq
\left(\begin{matrix}
0&h&\dots&h\\
h&\ddots&\ddots&\vdots
 \\\vdots&\ddots&\ddots&h \\h&\dots &h&0
\end{matrix}\right)\,\,\mbox{ with }\, \,h:=\beta \bar t(1-\alpha)(a- \sum_{i \in \mathcal{N}} \check d_{k,\ell}^i),\,k,\ell\in\mathcal{N},\,k\neq\ell\,.
\end{align}
Using a Taylor expansion, one can prove that $\exp(H)$ is of the form
\begin{align*}
\exp(H)\defeq
\left(\begin{matrix}
\chi(h)&\chi(h)-\exp(-h)&\dots&\chi(h)-\exp(-h)\\
\chi(h)-\exp(-h)&\ddots&\ddots&\vdots
 \\\vdots&\ddots&\ddots&\chi(h)-\exp(-h) \\\chi(h)-\exp(-h)&\dots &\chi(h)-\exp(-h)&\chi(h)
\end{matrix}\right)\,,
\end{align*}
where
$\chi(h):=1+\sum_{k\ge 1}{u_k}/{k!}$, with $(u_k)_{k \in \N\setminus\{0\}}$ satisfying the following recursive system\footnote{This system has a closed-form solution that can be determined by elementary methods. However, its expression is too inconvenient to report it in full length here.}
$$\begin{cases}
u_1=0\mbox{ and }
v_1=h\\
u_k=h(N-1)v_{k-1}\mbox{ and }
v_k=h\left[(n-2)v_{k-1}+u_{k-1}\right]\mbox{ for $k\ge 2$}
\end{cases}\,.$$
As $\chi(h) > \chi(h)-\exp(-h)$, it follows from Equation \eqref{eq derb} that for all distinct elements $i,k,\ell \in \mathcal{N},$ one has
$$\dfrac{\partial U_i(\cdot,\check D^{-i})} {\partial d^i_{\{i,\ell\}}}(\check D^i) >  \dfrac{\partial U_k(\cdot,\check D^{-k})} {\partial d^k_{\{i,\ell\}}}(\check D^{k})\,.$$
Using Proposition \ref{prop NEGlobal},  this yields the following characterisation:
 \begin{enumerate}[label=(\arabic*)]
   \setlength{\itemsep}{0pt}
\setlength{\parskip}{0pt}
\setlength{\parsep}{0pt}
\item The symmetric equilibrium is such that $h=0,$ or equivalently $\sum_{i\in\mathcal{N}} \check D^i=A,$ leading to $\chi(0)=1$, if only if
 $\rho  \le \delta\beta\bar t\alpha.$
\item The symmetric equilibrium is such that $h=\beta \bar t(1-\alpha)a,$ or equivalently $\sum_{i\in\mathcal{N}} \check D^i=O$  if and only if  $\rho  \ge \delta\beta\bar t\alpha \tilde\chi(\beta \bar t(1-\alpha)a)$ where $\tilde \chi(\beta \bar t(1-\alpha)a):=2\chi(\beta \bar t(1-\alpha)a)-\exp(-\beta \bar t(1-\alpha)a).$
\item The symmetric equilibrium is  interior if and only if  ${\rho}/{(\delta\beta\bar t\alpha)} \in \left(1,\tilde\chi(\beta \bar t(1-\alpha)a)\right)$.
\end{enumerate}
In the third case, the equilibrium is a local Homogeneous Interior Equilibrium, while, in the first two cases, there exists an equilibrium in local strategies that is equivalent to $\check D$ in the sense of Proposition \ref{prop indeterminacy}.
\end{ex}

As stated below, in view of Proposition \ref{prop weakint}, Example \ref{ex completenet} induces a Full Investment Equilibrium or an Interior Equilibrium in each network that is sufficiently connected.

\begin{prop} Consider an $(\alpha,\delta)$-Symmetric Game where there exists $\varepsilon>0$ such that for all   $k,\ell\in\mathcal{N},\,k\neq \ell,$ $a_{k,\ell} \geq \varepsilon.$
Then, one has in both the local and global games:
 \begin{enumerate}[wide=0pt, label=(\arabic*)]
   \setlength{\itemsep}{0pt}
\setlength{\parskip}{0pt}
\setlength{\parsep}{0pt}
\item If ${\rho}/{(\delta\beta\bar t\alpha)}  <\tilde\chi(\beta \bar t(1-\alpha)\varepsilon),$ there exists a Full Investment Equilibrium or an Interior Equilibrium.
\item If, moreover ${\rho}/{(\delta\beta\bar t\alpha)}>1,$ there exists an  Interior Equilibrium.
\end{enumerate}
\end{prop}

\subsection{Social efficiency and the PoA}

Using Lemma \ref{le marginalbis}, one can  provide a differential characterisation of Social Optima, as reported in the following proposition.

\begin{prop}\label{prop socgame}
A strategy profile $\hat D\in\mathcal{D}(A)$ is a Social Optimum if and only if for all $\{k,\ell\} \in\mathcal{E},$  one of the following alternative holds:
\small
 \begin{enumerate}[label=(\alph*)]
   \setlength{\itemsep}{0pt}
\setlength{\parskip}{0pt}
\setlength{\parsep}{0pt}
    \item $\rho<\sum_{i\in\mathcal{N}}\delta_i\left(  C_{i,k}(A,\hat D,\beta,\bar t,\x_0) x_{0_\ell} +C_{i,\ell}(A,\hat D,\beta,\bar t,\x_0) x_{0_k}  \right)$ and $\hat d_{\{k,\ell\}}=a_{\{k,l\}},$
    \item $\rho>\sum_{i\in\mathcal{N}}\delta_i\left(  C_{i,k}(A,\hat D,\beta,\bar t,\x_0) x_{0_\ell} +C_{i,\ell}(A,\hat D,\beta,\bar t,\x_0) x_{0_k}    \right)$and $\hat d_{\{k,\ell\}}=0,$
        \item $\rho=\sum_{i\in\mathcal{N}}\delta_i\left(  C_{i,k}(A,\hat D,\beta,\bar t,\x_0) x_{0_\ell} +C_{i,\ell}(A,\hat D,\beta,\bar t,\x_0) x_{0_k}    \right)$ and $ \hat d_{\{k,\ell\}} \in [0,a_{\{k,l\}}].$
\end{enumerate}
\end{prop}

Hence at a Social Optimum, there is investment in a link only if the sum of marginal utilities induced by the investment is larger than or equal to the investment cost. If the cost is smaller than the sum of marginal utilities, then the link is completely suppressed (case (a)). On the other hand, if the solution is interior, then the level of investment is such that  the sum of marginal utilities is exactly equal to the investment cost (case (c)). By analogy with the case of Nash Equilibria, we introduce the following definition.
\begin{dfn}[Social Optimum characterisation]\
\begin{itemize*}
\item A Full Investment Optimum is an optimum that satisfies, for all  $\{k,\ell\} \in\mathcal{E},$ case (a) of Proposition \ref{prop socgame}.
\item A No Investment Optimum is an optimum that satisfies, for all  $\{k,\ell\} \in\mathcal{E},$ case (b) of Proposition \ref{prop socgame}.
\item An Interior Optimum is an optimum that satisfies, for all  $\{k,\ell\} \in\mathcal{E},$ case (c) of Proposition \ref{prop socgame}.
\end{itemize*}
\end{dfn}
The comparison between Proposition $\ref{prop socgame}$ and Proposition \ref{prop NELocal} and  \ref{prop NEGlobal} underlines the fact that investment in contagion reduction has all the features of a public good problem. At equilibrium, the investment level is determined by the marginal utility of a single agent (the one with the largest willingness to pay) while social efficiency requires the investment level to be determined by the sum of all marginal utilities.  To quantify more precisely this inefficiency and its dependence on the network structure,  we shall use as a metric the PoA (see, e.g., \cite{papadimitriou2001algorithms,nisan2007algorithmic,omic2009protecting,hayel2014complete}). In our setting, where social welfare is always negative, the PoA  is defined as the ratio between the social welfare at a Nash Equilibrium and the welfare at a Social Optimum.  Accordingly, the PoA in the local and global games are respectively defined as follows
\begin{align}
\POAL&\defeq\frac{|\text{Worst social welfare at a local Nash Equilibrium}|}{|\text{Social welfare at a Social Optimum}|}\,, \label{eq defPoAL}
\\
\POAG&\defeq\frac{|\text{Worst social welfare at a global Nash Equilibrium}|}{|\text{Social welfare at a Social Optimum}|} \label{eq defPoAG}\,.
\end{align}
 By construction, the PoA is greater or equal to $1$ and equal to $1$ only when all Nash Equilibria of the game are socially optimal. An increasing PoA corresponds to an increasing social inefficiency of individual behaviours at a Nash Equilibrium.

Theorem  \ref{theo1} and \ref{theo2} below provide a partial characterisation of the PoA in the case of an $(\alpha)$-Homogeneous Game and Complete Network.

\begin{theo}\label{theo1}
Consider an $(\alpha)$-Homogeneous Game and a Complete Network. Assume that the global game $\mathcal{G}(\bdelta, A,\beta,\bar{t},\x_0,\rho)$ is such that: the worst Nash Equilibrium $\check D$ is not a Full Investment Equilibrium and one of the corresponding Social Optima $\hat D$ is not a Null Investment Optimum. Then
\begin{align*}
\POAG&\le\frac{\frac{N^2}{2\beta\bar t(1-\alpha)}+\mathbbm{1}^\top \sum_{i\in\mathcal{N}}\check D^i\mathbbm{1}}{\frac{N}{2\beta\bar t(1-\alpha)}+ \mathbbm{1}^\top \hat D\mathbbm{1}}\le N+\frac{2\beta\bar t(1-\alpha)}{N}\mathbbm{1}^\top A\mathbbm{1}\,.
\end{align*}
\end{theo}

The upper bound for $\POAL$ has a stronger dependence on the structure of the network.

\begin{theo}\label{theo2}
Consider an $(\alpha)$-Homogeneous Game and a Complete Network with $N\ge 3.$ Assume that the local game $\mathcal{L}(\bdelta, A,\beta,\bar{t},\x_0,\rho)$ is such that: the worst Nash Equilibrium $\bar D$ is not a Full Investment Equilibrium and one of the corresponding Social Optima $\hat D$ is not a Null Investment Optimum. Then
\begin{align*}
\POAL&\le\frac{\frac{1}{2\beta\bar t (1-\alpha)}\left[N^2\rho +\frac{(N-2)\alpha}{(N-1)}\sum_{i\in\mathcal{N}}K_i(\delta_i,A,\bar D,\bar D,\beta,\bar t,\x_0)\right]+\rho\mathbbm{1}^\top \sum_{i\in\mathcal{N}}\bar D^i\mathbbm{1}}{\rho\left[\frac{N}{2\beta\bar t(1-\alpha)}+ \mathbbm{1}^\top \hat D\mathbbm{1}\right]}
\\&\le N+\frac{(N-2)\alpha}{N(N-1)\rho}\sum_{i\in\mathcal{N}}K_i(\delta_i,A,O,A,\beta,\bar t,\x_0)+\frac{2\beta\bar t(1-\alpha)}{N}\mathbbm{1}^\top A\mathbbm{1}\,,
\end{align*}
where for any adjacency matrix $B$ and strategy profiles $D,D'$, $K_i(\delta_i,B,D,D',\beta,\bar t,\x_0):=\sum_{\substack{k\in\mathcal{N}\\k\neq i}}\delta_i(C_{i,k}(B,D,\beta,\bar t,\x_0)-C_{i,i}(B,D',\beta,\bar t,\x_0))$.

In particular, for all $i,k\in\mathcal{N},\,k\neq i$,
\begin{align*}
0\le\rho+\alpha\delta_i[C_{i,k}(A,\bar D,\beta,\bar t,\x_0)-C_{i,i}(A,\bar D,\beta,\bar t,\x_0)]\le 2[\rho-\alpha\delta_iC_{i,i}(A,\bar D,\beta,\bar t,\x_0)]\,.
\end{align*}
\end{theo}

Theorem \ref{theo1} and \ref{theo2} imply that the PoA grows at most linearly with the number of agents. As highlighted in the proposition below (that recalls the notations of Example \ref{ex completenet}), the case of a Symmetric Game shows that one cannot improve this linear bound.

\begin{prop}\label{prop exPOA}
Consider an $(\alpha,\delta)$-Symmetric Game and an $(a)$-Complete Network with $N\ge 3$.  Assume that ${\rho}/{ (\delta\beta\bar t\alpha)} \in \left(2,\tilde\chi(\beta \bar t(1-\alpha)a)\right),$ and let $\check D$ (resp. $\hat D$) be the worst Nash Equilibrium (resp. one of the corresponding Social Optima). Then
\begin{align*}\POAG=\POAL&=\frac{\frac{N^2\rho}{2\beta\bar t (1-\alpha)} -\frac{N(N-2)\delta\alpha\exp(-h)}{2(1-\alpha)}+\rho\mathbbm{1}^\top \sum_{i\in\mathcal{N}}\check D^i\mathbbm{1} }{\frac{N\rho}{2\beta\bar t(1-\alpha)}+\rho\mathbbm{1}^\top \hat D\mathbbm{1}}\nonumber\\&\le N- \frac{(N-2)\delta\beta \bar t\alpha\exp(- h)}{\rho}+\frac{2\beta\bar t(1-\alpha)}{N}\mathbbm{1}^\top A\mathbbm{1}\,,\nonumber\end{align*}
where $h\defeq \beta\bar t(1-\alpha)(a-\sum_{i\in\mathcal{N}}\check d^i_{k,\ell})> 0$, for any $k,\ell\in\mathcal{N},\,k\neq \ell$. In particular, $\rho -\delta\beta\bar t\alpha\exp(-h)\ge 0.$
\end{prop}

Our previous results highlight the fact that individual strategic behaviours can lead to major inefficiencies in the containment of epidemic spreading. In particular, there can be complete free-riding of other players on the investment of the agent that is the most affected by the epidemic (Proposition \ref{prop NELocal} and \ref{prop NEGlobal}) and the inefficiency can scale up to linearly with the number of agents (Theorem \ref{theo1} and \ref{theo2} and Proposition \ref{prop exPOA}). In other words, individual strategic behaviours can be highly inefficient in terms of social welfare as soon as there is a large number of agents involved. This is the case in real-world applications whether one considers epidemic spreading between individuals at the domestic scale or between countries at the global scale.

Against this backdrop, it is natural to search for a public policy response for the prevention of epidemic spreading. During the recent COVID-19 outbreak, a widespread policy response
has been the implementation of social distancing measures that have reduced, in a uniform way, the scale of social interactions. Formally, we can define the social distancing policy at level $\kappa \in \mathbb{R}_+$ as restricting social interactions to $C(A,\kappa):=(c_{i,j})_{i,j \in \mathcal{N}}$ such that for all $i,j \in \mathcal{N},$ $c_{i,j}:= \kappa {a_{i,j}}/{\sum_{k \in \mathcal{N}} a_{i,k}}$ (or equivalently $c_{i,j}:=  \kappa {a_{i,j}}/{\sum_{k \in \mathcal{N}} a_{k,i}}$  as $A$ is symmetric). The level $\kappa \in \mathbb{R}_+$ must be such that $A-C(A,\kappa)\ge O$.  Hence, the social distancing policy amounts  to bounding the level of social interactions of each agent to a fixed level. In practice, this has been implemented by massive restrictions on socio-economic activities such as  interdiction of public gatherings, closing of schools and businesses, and travel restrictions. A formal analysis of this policy in our framework shows it can be socially efficient, at least if the initial contagion probability and the disutility are assumed to be uniform. Namely, it is optimal in the following sense.

\begin{prop} \label{prop dist} Consider a $(\alpha,\delta)$-Symmetric Game and assume that $\sum_{k\in\mathcal{N}} a_{i,k}>0$ for all $i\in\mathcal{N}$.
\begin{enumerate}
   \setlength{\itemsep}{0pt}
\setlength{\parskip}{0pt}
\setlength{\parsep}{0pt}
 \item If $2\delta\beta \bar t \alpha\ge \rho,$ then $\kappa=0$ is optimal and the optimal social distancing measure involves the suppression of every link.
\item If $2\delta\beta \bar t \alpha< \rho,$ then for every $\varepsilon>0,$ there exists $\bar T>0$ such
that for $\bar t\ge \bar T$, one can find $2\delta\beta \bar t \alpha< \rho\le2\delta\beta \bar t \alpha\exp (\beta \bar t(1-\alpha)  \times\min_{i\in\mathcal{N}}\sum_{k\in\mathcal{N}} a_{i,k})$ for which there exists an admissible $\kappa >0$ satisfying
\begin{equation*} \hat \Pi  (A-C(A,\kappa)) \geq \max_{D \in \mathcal{D}(A)}  \hat \Pi (D)- \varepsilon\,. \end{equation*}
In this case, for every $\varepsilon>0$ and for $\bar t$ large enough, there is an $\varepsilon-$optimal social distancing measure which does not involve any suppression of link.
\end{enumerate}
\end{prop}
Hence, uniform reduction of social interactions appears as being an extremely efficient policy in our framework. This appears as a natural counterpart to existing results  in the literature that emphasise the role of highly connected nodes, e.g.,``super spreaders", in epidemic propagation (see, e.g., \cite{pastor2001epidemic,pastor2015epidemic}). Our result would benefit from further research on the extent to which one could relax the assumption that the ``implicit" costs of connectivity reduction are homogeneous. This assumption indeed  neglects the fact that certain actors might value more social interactions because of their  economic, psychological, or social characteristics. Moreover, results on the determination of the optimal level of social distancing would also be welcome. In practice, the level of social distancing has been determined according to policy decisions about the socially/economically acceptable rate of contagion, rather than inferred from individual preferences.

\section{The PoK}\label{se pok}

Social distancing measures can be implemented at the domestic scale in order to reduce the propagation of epidemics between individuals. However, at the international scale, there is no authority entitled to implement such coercive measures. Furthermore, individual countries can take measures to reduce their interactions with other countries, e.g., border closures, but cannot directly reduce interactions between two other countries. They are thus, by default,  in the framework of a local game. One could nevertheless consider schemes in which countries with a higher disutility from infection subsidise investments in other parts of the network to reduce global contagiousness. This would turn the problem into a global game. In order to compare outcomes in these two situations, we introduce the notion of PoK, which corresponds to the ratio between the social welfare at the worst  equilibrium of the local game and at the best equilibrium of the global game.
\begin{align*}
\POK\defeq\frac{|\text{Worst social welfare at a Nash Equilibrium of the local game}|}{|\text{Best social welfare at a Nash Equilibrium of the global game}|}\,.
\end{align*}
In particular, if $\POK > 1$, then global equilibria are better than local ones. On the contrary if global and local equilibria coincide (as in Example  \ref{ex completenet}),
$\POK= 1$. The latter implies in particular  that an increase in the set of admissible strategies does not necessarily induce an increase in individual welfare. Sometimes it could even lead to more free riding.
In general, the value of the $\POK$ is determined by the network structure and the  individual disutilities  associated to contagion,  measured by the coefficients $\delta_i,\,i\in\mathcal{N}$.  In particular, following the lines of the proofs of Theorem \ref{theo1}-\ref{theo2}, one can provide an explicit lower bound on the  $\POK$, as detailed in the theorem below.

\begin{theo}\label{theo3}
Consider an $(\alpha)$-Homogeneous Game and a Complete Network with $N\ge 3$. Assume that the local game $\mathcal{L}(\bdelta, A,\beta,\bar{t},\x_0,\rho)$ and global game $\mathcal{G}(\bdelta, A,\beta,\bar{t},\x_0,\rho)$ are such that: the worst local Nash Equilibrium $\bar D$ is an Homogeneous Interior Equilibrium and the best global Nash Equilibrium $\check D$ is not a Full Investment Equilibrium. Then
\begin{align}
\POK&\ge\frac{\frac{1}{\beta\bar t (1-\alpha)}\left[N(N-1)\rho-(N-2)\alpha\sum_{i\in\mathcal{N}}\delta_iC_{i,i}(A,\bar D,\beta,\bar t,\x_0)\right]+\rho\mathbbm{1}^\top \sum_{i\in\mathcal{N}}\bar D^i\mathbbm{1}}{\rho\left[\frac{N^2}{2\beta\bar t (1-\alpha)}+\mathbbm{1}^\top \sum_{i\in\mathcal{N}}\check D^i\mathbbm{1}\right]}\label{eq POKor}\\&\ge \frac{\frac{N}{\beta\bar t (1-\alpha)} }{\frac{N^2}{2\beta\bar t (1-\alpha)}+\mathbbm{1}^\top A\mathbbm{1}}\,.\nonumber
\end{align}
In particular, for all $i,k\in\mathcal{N},\,k\neq i$,
\begin{align}\label{eq POK+}
0\le\rho+\alpha\delta_i[C_{i,k}(A,\bar D,\beta,\bar t,\x_0)-C_{i,i}(A,\bar D,\beta,\bar t,\x_0)]= 2[\rho-\alpha\delta_iC_{i,i}(A,\bar D,\beta,\bar t,\x_0)]\,.
\end{align}
 \end{theo}

As hinted above, in the case of a Complete Network, under the assumptions of  Proposition \ref{prop exPOA}, the conditions of Theorem \ref{theo3} are satisfied and $\POK=1$.  Moreover, Equation \eqref{eq POKor}-\eqref{eq POK+} above highlight the fact that the PoK increases when there exists  agents $i\in\mathcal{N}$ with a large disutility of contagion $\delta_i$ that are highly connected to other nodes in the network, as measured by $C_{i,k}(\cdot),\,i,k\in\mathcal{N},\,k\neq i.$ A salient example of such case is that of an $(a)$-Complete Network where one of the agents has a much higher disutility of contagion than its peers. We thus intend to study this example and more specifically to consider the limit case where $\bm{\delta}=  \bm\delta \e^i$  for some $i\in\mathcal{N},$ i.e. where the disutility of all other agents is negligible with respect to that of agent $i.$ In this setting there is no indeterminacy on the agent that is playing and we can give a necessary condition for global strategies to dominate local ones. This is the aim of the following proposition.

\begin{prop}\label{prop exPOK}
Fix $i\in\mathcal{N}$. Consider the local game $\mathcal{L}(\bm\delta \e^i, A,\beta,\bar{t},\x_0,\rho)$ for an $(\alpha)$-Homogeneous Game and an $(a)$-Complete Network with $N\ge 3$. If
$$\delta_i\beta\bar t\alpha<\rho<\delta_i\beta\bar t\alpha\left(\frac{\sinh(\sqrt{N-1}\beta\bar t(1-\alpha)a)}{\sqrt{N-1}}+\cosh(\sqrt{N-1}\beta\bar t(1-\alpha)a)\right)\,,$$
then one has:
 \begin{enumerate}[label=(\arabic*)]%
   \setlength{\itemsep}{0pt}
\setlength{\parskip}{0pt}
\setlength{\parsep}{0pt}
\item Any local equilibrium  $\bar D$ is an Homogeneous Interior Equilibrium and is such that $\bar D^j=O$, for all $j\in\mathcal{N},\,j\neq i,$ and $\bar d^i_{i,k}=\bar d^i_{k,i}=h$ for all $\{i,k\} \in \mathcal{E}$ for some $h\in [0,a]$.

\item Global strategies dominate local ones if and only if the parameters $\beta,\bar t,\alpha,$ and $a$ are such that
\begin{align*}
\sinh(\sqrt{N-1}\beta\bar t(1-\alpha)(a-h))>\sqrt{N-1}\cosh(\sqrt{N-1}\beta\bar t(1-\alpha)(a-h))\,.
\end{align*}
 \end{enumerate}
\end{prop}

\section{Conclusion}
\label{se conc}
In this paper, we have investigated the prophylaxis of epidemic spreading from a normative point of view in a game-theoretic setting.  Agents have the common objective to reduce the speed of  propagation of an epidemic of the SI type through investments in the reduction of the contagiousness of network links.  Despite this common objective, strategic behaviours and free-riding can lead to major inefficiencies. We have shown that the PoA can scale up to linearly in our setting. This strongly calls for public intervention to reduce the speed of diffusion. In this respect, we have shown that  a policy of uniform reduction of social interactions, akin to the social distancing measures enforced during the COVID-19 pandemic, can be $\varepsilon$-optimal in a wide range of networks. Such policies thus have strong normative foundations. Our results however assume that the cost of reducing interactions is uniform among agents. The validity of this assumption strongly depends on the scope of the analysis: it is a much more benign approximation when the focus is on public health than in the case where economic and financial considerations ought to be taken into account.

We have partly accounted for heterogeneity as far as the benefits of prophylaxis are concerned. In this setting, we have shown that allowing agents to subsidise investments in the reduction of contagiousness in distant parts of the network can be Pareto improving. This  result calls for further research on the design of mechanisms to improve the efficiency of cooperation against epidemic spreading.

 \renewcommand{\theequation}{A-\arabic{equation}}
  \setcounter{equation}{0}
  \section*{Appendix}
\begin{proof}[{\bf of Lemma \ref{le eigenvec}}] The proof follows from the spectral decomposition of $A-\sum_{i \in \mathcal{N}} D^i$ and a direct application of the Perron-Frobenius theorem. See \citet[Appendix C]{lee2019transient} for details.
\end{proof}
\begin{proof}[{\bf of Lemma \ref{le marginal}}]
Let $i\in\mathcal{N}$ and $(D^i, D^{-i})$ be a strategy profile in $\mathcal{S}(A)$. For $k,\ell\in\mathcal{N}$, one has
\begin{align*}\dfrac{\partial U_i(\cdot, D^{-i})} {\partial d^i_{\{k,\ell\}}}(D^i)&=   \delta_i  <\textbf{e}^i,\beta \bar t\exp{(\beta \bar t(A-\sum_{j\in\mathcal{N},\,j \not = i}  D^j- D^i)\diago(1-\x_0))} (\hat{I}^{k,\ell}+ \hat{I}^{\ell,k})\x_0> \,,
\end{align*}
where for any $k,j\in\mathcal{N}$, $\hat{I}^{k,j}$ is the $N-$ dimensional square matrix with null entries except on the $k^{\text{th}}-$row and $j^{\text{th}}-$column for which the entry is equal to one. This leads to Equation \eqref{eq der}.
Moreover, for all $k,\ell,p,q\in\mathcal{N}$, we compute
\begin{align*}
&\dfrac{\partial U_i(\cdot, D^{-i})}{\partial d^i_{k,\ell}\partial d^i_{p,q}}(D^i)\\&=-<\textbf{e}^i,(\beta \bar t)^2\exp{(\beta \bar t(A-\sum_{j\in\mathcal{N},\,j \not = i}  D^j- D^i)\text{diag}(1-\x_0))}\hat{I}^{p,q}\text{diag}(1-\x_0)\hat{I}^{k,\ell}\x_0>\\&
=\begin{cases}
-(\beta \bar t)^2\left(\exp{(\beta \bar t(A-\sum_{j\in\mathcal{N},\,j \not = i}  D^j- D^i)\text{diag}(1-\x_0))}\right)_{i,p}(1-x_{0_q})x_{0_\ell}\mbox{ if } k=q\\
0\mbox{ otherwise}
\end{cases}\,.
\end{align*}
Therefore the Hessian matrix of $U_i(\cdot, D^{-i})$ on $\mathcal{S}_i( A,D^{-i})$ is the matrix of a quadratic form that is negative semi-definite. The concavity property thus follows.
\end{proof}
\begin{proof}[{\bf of Theorem \ref{th NEEx}}]
We know from Remark \ref{re setprop} that the sets $\mathcal{S}(A)$ and $\mathcal{K}(A)$ of admissible strategies are compact and convex, and from Lemma \ref{le marginal} that the objective function $\Pi$ is concave on $\mathcal{S}_i(A, D^{-i})$ and  $\mathcal{K}_i(A, D^{-i})$ with $i\in\mathcal{N}$ and $D^{-i} \in (\mathbb{S}^N(\R_+))^{N-1}$. Moreover, since $\Pi$ is continuous in its arguments, we therefore conclude from \citet[Theorem 1]{rosen1964existence} that a Nash Equilibrium exists.
\end{proof}
\begin{proof}[{\bf of proposition \ref{prop NELocal}}]
For $1\le i\le N$, the optimisation programme for characterising the Nash Equilibria writes
{
\begin{align}
&\max_{D^i\in \mathbb{M}^{N}} \Pi_i(D^i,\bar D^{-i})\nonumber\\
&\mbox{ subject to: }\nonumber\\
& d^i_{k,\ell}=d^i_{\ell,k},\,\forall\,k,\ell\in \mathcal{N},\,\ell<k\,,\label{eq cond1}\\
&d^i_{k,\ell}=0,\,\forall\,k,\ell\in\mathcal{N},\,\ell> k,\,k\mbox{ and }\ell\neq i\,,\nonumber\\
&d^i_{k,\ell}\ge 0,\,\forall\,k,\ell\in \mathcal{N},\,\ell> k,\,k\mbox{ or }\ell= i\,,\nonumber\\
&(A-\bar D^{-i}-D^i)_{k,\ell}\ge 0,\,\forall\,k,\ell\in \mathcal{N},\,\ell\ge k\,.\label{eq cond4}
\end{align}}
\normalsize
Applying the Karush-Kuhn-Tucker conditions, we obtain that $\bar D^i\in\mathcal{K}_i(A,\bar D^{-i})$ is a solution if and only if, for any $\{k,\ell\}\in\mathcal{E}$,  one of the following cases holds, assuming without loss of generality that ${\partial U_k(\cdot,\bar D^{-k})}/{\partial d^k_{\{k,\ell\}}}(\bar D^k)\le {\partial U_\ell(\cdot,\bar D^{-\ell})}/ {\partial d^\ell_{\{k,\ell\}}}(\bar D^\ell)$:
 \begin{enumerate}[wide=0pt,label=(\alph*)]
   \setlength{\itemsep}{0pt}
\setlength{\parskip}{0pt}
\setlength{\parsep}{0pt}
\item  \small
 \begin{enumerate}[wide=0pt, label=(\roman*)]
   \setlength{\itemsep}{0pt}
\setlength{\parskip}{0pt}
\setlength{\parsep}{0pt}
\item $\rho< {\partial U_k(\cdot,\bar D^{-k})}/ {\partial d^k_{\{k,\ell\}}}(\bar D^k)$ and $\bar d^k_{\{k,\ell\}}+\bar d^\ell_{\{k,\ell\}}=a_{\{k,\ell\}},$
\item ${\partial U_k(\cdot,\bar D^{-k})} /{\partial d^k_{\{k,\ell\}}}(\bar D^k)<\rho<{\partial U_\ell(\cdot,\bar D^{-\ell})}/ {\partial d^\ell_{\{k,\ell\}}}(\bar D^\ell)$ and $\bar d^k_{\{k,\ell\}}=0$, $\bar d^\ell_{\{k,\ell\}}=a_{\{k,\ell\}},$
\item ${\partial U_k(\cdot,\bar D^{-k})} /{\partial d^k_{\{k,\ell\}}}(\bar D^k)=\rho<{\partial U_\ell(\cdot,\bar D^{-\ell})}/ {\partial d^\ell_{\{k,\ell\}}}(\bar D^\ell)$ and $\bar d^k_{\{k,\ell\}}\in[0,a_{\{k,\ell\}}]$, $\bar d^\ell_{\{k,\ell\}}=a_{\{k,\ell\}}-\bar d^k_{\{k,\ell\}},$
\end{enumerate}
\item $\rho> {\partial U_\ell(\cdot,\bar D^{-\ell})} /{\partial d^\ell_{\{k,\ell\}}}(\bar D^\ell)$ and $\bar d^k_{\{k,\ell\}}=\bar d^\ell_{\{k,\ell\}}=0,$
\item
 \begin{enumerate}[wide=0pt, label=(\roman*)]
   \setlength{\itemsep}{0pt}
\setlength{\parskip}{0pt}
\setlength{\parsep}{0pt}
\item ${\partial U_k(\cdot,\bar D^{-k})}/ {\partial d^k_{\{k,\ell\}}}(\bar D^k)<\rho={\partial U_\ell(\cdot,\bar D^{-\ell})} /{\partial d^\ell_{\{k,\ell\}}}(\bar D^\ell)$ and $\bar d^k_{\{k,\ell\}}=0$, $\bar d^\ell_{\{k,\ell\}}\in[0,a_{\{k,\ell\}}],$
\item ${\partial U_k(\cdot,\bar D^{-k})} /{\partial d^k_{\{k,\ell\}}}(\bar D^k)=\rho={\partial U_\ell(\cdot,\bar D^{-\ell})}/ {\partial d^\ell_{\{k,\ell\}}}(\bar D^\ell)$ and $0\le \bar d^k_{\{k,\ell\}}+\bar d^\ell_{\{k,\ell\}}\le a_{\{k,\ell\}}.$
\end{enumerate}
\normalsize
\end{enumerate}
\end{proof}
\begin{proof}[{\bf of Proposition \ref{prop NEGlobal}}] For $1\le i\le N$, the optimisation programme for characterising the Nash Equilibria writes
{
\begin{align*}
&\max_{D^i\in \mathbb{M}^{N}} \Pi_i(D^i,\check D^{-i})\nonumber\\
&\mbox{ subject to: }\\
&\mbox{Condition \eqref{eq cond1}-\eqref{eq cond4}}\,,\nonumber\\
&d^i_{k,\ell}\ge 0,\,\forall\,k,\ell\in \mathcal{N},\,\ell> k\,.
\end{align*}}
\normalsize
Applying the Karush-Kuhn-Tucker conditions, we obtain that $\check D^i\in\mathcal{S}_i(A,\check D^{-i})$ is a solution if and only if, for any $\{k,\ell\}\in\mathcal{E}$,  one of the following cases holds:
 \begin{enumerate}[wide=0pt, label=(\alph*)]
   \setlength{\itemsep}{0pt}
\setlength{\parskip}{0pt}
\setlength{\parsep}{0pt}
\item  \small
 \begin{enumerate}[wide=0pt,label=(\roman*)]
   \setlength{\itemsep}{0pt}
\setlength{\parskip}{0pt}
\setlength{\parsep}{0pt}
\item $\rho<\min_{i\in\mathcal{N}}\left({\partial U_i(\cdot,\check D^{-i})}/ {\partial d^i_{\{k,\ell\}}}(\check D^i)\right)$ and $\sum_{q\in\mathcal{N}}\check d^q_{\{k,\ell\}}=a_{\{k,\ell\}},$
\item there exists $i,j\in\mathcal{N},$ such that ${\partial U_i(\cdot,\check D^{-i})}/ {\partial d^i_{\{k,\ell\}}}(\check D^i)<\rho<{\partial U_j(\cdot,\check D^{-j})}/ {\partial d^j_{\{k,\ell\}}}(\check D^j)$
and $\check d^q_{\{k,\ell\}}=0$ for all $q\in\mathcal{A}^i(\check D)$, $\sum_{q\in\bar{\mathcal{A}}^j(\check D)}\check d^q_{\{k,\ell\}}=a_{\{k,\ell\}},$ where for a given $h\ge 1$,
\begin{gather*}
\mathcal{A}^h(\check D)\defeq\left\{1\le r\le N:\,\dfrac{\partial U_r(\cdot,\check D^{-r})} {\partial d^r_{\{k,\ell\}}}(\check D^r)\le \dfrac{\partial U_h(\cdot,\check D^{-h})} {\partial d^h_{\{k,\ell\}}}(\check D^h)\right\}\,,
\\\mbox{ and }\\\bar{\mathcal{A}}^h(\check D)\defeq\left\{1\le r\le N:\,\dfrac{\partial U_r(\cdot,\check D^{-r})} {\partial d^r_{\{k,\ell\}}}(\check D^r)\ge \dfrac{\partial U_h(\cdot,\check D^{-h})} {\partial d^h_{\{k,\ell\}}}(\check D^h)\right\}\,,
\end{gather*}
\item there exists $i,j\in\mathcal{N},$ such that ${\partial U_i(\cdot,\check D^{-i})}/ {\partial d^i_{\{k,\ell\}}}(\check D^i)=\rho<{\partial U_j(\cdot,\check D^{-j})}/ {\partial d^j_{\{k,\ell\}}}(\check D^j)$ and
$0\le \sum_{q\in\hat{\mathcal{A}}^i(\check D)}\check d^q_{\{k,\ell\}}\le a_{\{k,\ell\}},$ $\check d^q_{\{k,\ell\}}=0$ for all $q\in\left(\bar{\mathcal{A}}^i(\check D)\right)^\text{c}$, $\sum_{q\in\bar{\mathcal{A}}^j(\check D)}\check d^q_{\{k,\ell\}}=a_{\{k,\ell\}}-\sum_{q\in\hat{\mathcal{A}}^i(\check D)}\check d^q_{\{k,\ell\}}$, where for a given $h\ge 1$,
 \begin{align*}\hat{\mathcal{A}}^h(\check D)\defeq\left\{1\le r\le N:\,\dfrac{\partial U_r(\cdot,\check D^{-r})} {\partial d^r_{\{k,\ell\}}}(\check D^r)= \dfrac{\partial U_h(\cdot,\check D^{-h})} {\partial d^h_{\{k,\ell\}}}(\check D^h)\right\}\,,
\end{align*}
\end{enumerate}
\item $\rho>\max_{i\in\mathcal{N}}\left({\partial U_i(\cdot,\check D^{-i})} /{\partial d^i_{\{k,\ell\}}}(\check D^i)\right)$ and $\sum_{q\in\mathcal{N}}\check d^q_{\{k,\ell\}}=0,$
\item
there exists $j\in\mathcal{N},$ such that $\rho={\partial U_j(\cdot,\check D^{-j})}/ {\partial d^j_{\{k,\ell\}}}(\check D^j)$
and $\check d^q_{\{k,\ell\}}=0$ for all $q\in\left(\bar{\mathcal{A}}^j(\check D)\right)^\text{c}$, $0\le \sum_{q\in\hat{\mathcal{A}}^j(\check D)}\check d^q_{\{k,\ell\}}\le a_{\{k,\ell\}}.$
\end{enumerate}
\end{proof}
\begin{proof}[{\bf of Proposition \ref{prop socgame}}]
For $1\le i\le N$, the optimisation programme for characterising the Social Optima writes
{
\begin{align}
&\max_{D\in \mathbb{M}^{N}} \hat\Pi(D)\nonumber\\
&\mbox{ subject to: }\nonumber\\
& d_{k,\ell}=d_{\ell,k},\,\forall\,k,\ell\in \mathcal{N},\,\ell<k\,,\nonumber\\
&d_{k,\ell}\ge 0,\,\forall\,k,\ell\in \mathcal{N},\,\ell> k\,,\nonumber\\
&(A- D)_{k,\ell}\ge 0,\,\forall\,k,\ell\in \mathcal{N},\,\ell\ge k\,.\nonumber
\end{align}}
\normalsize
The proof is then a straightforward adaptation of the proof of Proposition \ref{prop NELocal} and \ref{prop NEGlobal}.
\end{proof}
\begin{proof}[{\bf of Theorem \ref{theo1}}]
It follows from the assumption on $\check D$ that for all $i\in\mathcal{N}$,
\begin{align}\label{eq eqint0}
\sum_{\substack{k,\ell\in\mathcal{N}\\k\neq\ell}}\dfrac{\partial U_i(\cdot,\check D^{-i})} {\partial d^i_{\{k,\ell\}}}(\check D^i)\le N(N-1)\rho\,.
\end{align}
Therefore appealing to Equation \eqref{eq derveq} and Equation \eqref{eq eqint0}, we obtain
\begin{align}
\sum_{i\in\mathcal{N}}\Pi_i(\check D^i,\check D^{-i})&=\sum_{i\in\mathcal{N}}U_i(\check D^i,\check D^{-i})-\rho\mathbbm{1}^\top \sum_{i\in\mathcal{N}}\check D^i\mathbbm{1}\nonumber\\
&=-\frac{1}{2(N-1)\beta\bar t(1-\alpha)}\sum_{i\in\mathcal{N}}\sum_{\substack{k,\ell\in\mathcal{N}\\k\neq\ell}}\dfrac{\partial U_i(\cdot,\check D^{-i})} {\partial d^i_{\{k,\ell\}}}(\check D^i)-\rho \mathbbm{1}^\top \sum_{i\in\mathcal{N}}\check D^i\mathbbm{1}\nonumber\\
&\ge-\rho\left[\frac{N^2}{2\beta\bar t(1-\alpha)}+\mathbbm{1}^\top \sum_{i\in\mathcal{N}}\check D^i\mathbbm{1}\right]\,.\label{eq WNE}
\end{align}
Similarly, it follows from the assumption on $\hat D$ that it is such that for all $k,\ell\in\mathcal{N},\,k\neq\ell$,
\begin{align}\label{eq eqint2}
\sum_{i\in\mathcal{N}}\dfrac{\partial \hat v_i(\hat D)} {\partial d_{\{k,\ell\}}}\ge\rho\,.
\end{align}
We deduce from Equation \eqref{eq derveq2} and Equation \eqref{eq eqint2},
\begin{align}
\hat\Pi(\hat D)&=\sum_{i\in\mathcal{N}}\hat v_i(\hat D)-\rho \mathbbm{1}^\top \hat D\mathbbm{1}\nonumber\\
&=-\frac{1}{2(N-1)\beta\bar t(1-\alpha)}\sum_{i\in\mathcal{N}}\sum_{\substack{k,\ell\in\mathcal{N}\\k\neq\ell}}\dfrac{\partial \hat v_i(\hat D)} {\partial d_{\{k,\ell\}}}-\rho \mathbbm{1}^\top \hat D\mathbbm{1}\nonumber\\
&\le-\rho\left[\frac{N}{2\beta\bar t(1-\alpha)}+\mathbbm{1}^\top \hat D\mathbbm{1}\right]\,.\label{eq WSO}
\end{align}
Combining Equation \eqref{eq WNE} and \eqref{eq WSO} and using Equation \eqref{eq defPoAG}, we obtain
the result.
\end{proof}
\begin{proof}[{\bf of Theorem \ref{theo2}}]
We know from Equation \eqref{eq der} that for all $i,k,\ell\in\mathcal{N}$,
\begin{equation}\label{eq PoAL1}
\dfrac{\partial U_i(\cdot,\bar D^{-i})} {\partial d^i_{\{k,\ell\}}}(\bar D^i)=\alpha\delta_i[C_{i,k}(A,\bar D,\beta,\bar t,\x_0)+C_{i,\ell}(A,\bar D,\beta,\bar t,\x_0)]\,.
\end{equation}
Therefore, it follows from the assumption on $\bar D$ that for all $i,\ell\in\mathcal{N},\,i\neq\ell$,
\begin{align}\label{eq PoAL1ter}
\dfrac{\partial U_i(\cdot,\bar D^{-i})} {\partial d^i_{\{i,\ell\}}}(\bar D^i)=\alpha\delta_i[C_{i,i}(A,\bar D,\beta,\bar t,\x_0)+C_{i,\ell}(A,\bar D,\beta,\bar t,\x_0)]\le \rho\,.
\end{align}
Hence Equation \eqref{eq PoAL1}-\eqref{eq PoAL1ter} give that for all distinct $i,k,\ell\in\mathcal{N}$,
\begin{align}
\dfrac{\partial U_i(\cdot,\bar D^{-i})} {\partial d^i_{\{k,\ell\}}}(\bar D^i)&\le \rho+\alpha\delta_i[C_{i,k}(A,\bar D,\beta,\bar t,\x_0)-C_{i,i}(A,\bar D,\beta,\bar t,\x_0)]\label{eq PoAL3a}\\
&\le 2[\rho-\alpha\delta_iC_{i,i}(A,\bar D,\beta,\bar t,\x_0)]\,.\label{eq PoAL3b}
\end{align}
We deduce from Equation \eqref{eq PoAL3a}-\eqref{eq PoAL3b} that for all $i\in\mathcal{N}$,
\small
\begin{align}
\sum_{\substack{k,\ell\in\mathcal{N}\\k\neq \ell}}\dfrac{\partial U_i(\cdot,\bar D^{-i})} {\partial d^i_{\{k,\ell\}}}(\bar D^i)
&=\sum_{\substack{\ell\in\mathcal{N}\\\ell\neq i}}\dfrac{\partial U_i(\cdot,\bar D^{-i})} {\partial d^i_{\{i,\ell\}}}(\bar D^i)+\sum_{\substack{k\in\mathcal{N}\\k\neq i}}\dfrac{\partial U_i(\cdot,\bar D^{-i})} {\partial d^i_{\{k,i\}}}(\bar D^i)+\sum_{\substack{k,\ell\in\mathcal{N}\\k\neq \ell\\k,\ell\neq i}}\dfrac{\partial U_i(\cdot,\bar D^{-i})} {\partial d^i_{\{k,\ell\}}}(\bar D^i)
\nonumber\\&\le N(N-1)\rho+(N-2)\alpha\delta_i\sum_{k\in\mathcal{N},\,k\neq i}(C_{i,k}(\cdot)-C_{i,i}(\cdot))(A,\bar D,\beta,\bar t,\x_0)\label{eq PoAL7a}
\\&\le2(N-1)[(N-1)\rho-(N-2)\alpha\delta_iC_{i,i}(A,\bar D,\beta,\bar t,\x_0)]\,.\nonumber
\end{align}
\normalsize
In particular, we observe from the non-decreasing property of marginal utilities (recall Lemma \ref{le marginal}), that for all $i\in\mathcal{N}$,
\small
\begin{align*}
\rho+\alpha\delta_i[C_{i,k}(\cdot)-C_{i,i}(\cdot)](A,\bar D,\beta,\bar t,\x_0)\ge0\,\forall\,k\in\mathcal{N},\,k\neq i,
\mbox{ and }
\rho-\alpha\delta_iC_{i,i}(A,\bar D,\beta,\bar t,\x_0)\ge0\,.
\end{align*}
\normalsize
Finally, after appealing to Equation \eqref{eq derveq} and Equation \eqref{eq PoAL7a}, we obtain
\small
\begin{align*}
\hspace{-1em}\sum_{i\in\mathcal{N}}\Pi_i(\bar D^i,\bar D^{-i})&=\sum_{i\in\mathcal{N}}U_i(\bar D^i,\bar D^{-i})-\rho\mathbbm{1}^\top \sum_{i\in\mathcal{N}}\bar D^i\mathbbm{1}\nonumber\\
&=-\frac{1}{2(N-1)\beta\bar t(1-\alpha)}\sum_{i\in\mathcal{N}}\sum_{\substack{k,\ell\in\mathcal{N}\\k\neq\ell}}\dfrac{\partial U_i(\cdot,\bar D^{-i})} {\partial d^i_{\{k,\ell\}}}(\bar D^i)-\rho \mathbbm{1}^\top \sum_{i\in\mathcal{N}}\bar D^i\mathbbm{1}\nonumber\\
&\ge -\frac{1}{2\beta\bar t (1-\alpha)}\left[N^2\rho +\frac{(N-2)\alpha}{(N-1)}\sum_{i\in\mathcal{N}}K_i(\delta_i,A,\bar D,\bar D,\beta,\bar t,\x_0)\right]-\rho\mathbbm{1}^\top \sum_{i\in\mathcal{N}}\bar D^i\mathbbm{1}\,.
\end{align*}
\normalsize
Appealing to Equation \eqref{eq WSO} and Equation \eqref{eq defPoAL}, the result follows.
\end{proof}
\begin{proof}[{\bf of Proposition \ref{prop exPOA}}]
According to Example \ref{ex completenet}, under the holding assumptions, $\check D$ is a local Homogeneous Interior Equilibrium which yields an equilibrium network of the form given by Identity \eqref{eq matH}. More precisely, for all distinct $i,k,\ell\in\mathcal{N}$,
\begin{align*}
\dfrac{\partial U_i(\cdot,\check D^{-i})} {\partial d^i_{\{i,\ell\}}}(\check D^i)&=\delta\beta\bar t\alpha\left(2\chi(h)-\exp(-h)\right)=\rho\mbox{ and }\dfrac{\partial U_i(\cdot,\check D^{-i})} {\partial d^i_{\{k,\ell\}}}(\check D^i)=\rho-\delta\beta\bar t\alpha\exp(-h)\,.
\end{align*}
In particular, it follows from the non-decreasing property of marginal utilities (recall Lemma \ref{le marginal}) that $\rho-\delta\beta\bar t\alpha\exp(-h)\ge 0$. It is then straightforward to check that
\begin{equation*}
\sum_{i\in\mathcal{N}}\Pi_i(\check D^i,\check D^{-i})=-\frac{N^2\rho}{2\beta\bar t (1-\alpha)} +\frac{N(N-2)\delta\alpha\exp(-h)}{2(1-\alpha)}-\rho\mathbbm{1}^\top \sum_{i\in\mathcal{N}}\check D^i\mathbbm{1}\,.
\end{equation*}
On the other hand, one can assume without loss of generality that $\hat D$ is of the form
\begin{align*}
\hat D\defeq
\left(\begin{matrix}
0&\hat d&\dots&\hat d\\
\hat d&\ddots&\ddots&\vdots
 \\\vdots&\ddots&\ddots&\hat d \\\hat d&\dots &\hat d&0
\end{matrix}\right)\,,\mbox{ for some }\hat d\ge 0\,.
\end{align*}
Indeed, by concavity of $\hat \Pi,$ we know the set of Social Optima  is convex. Moreover, given  the symmetry of the game, the set of Social Optima shall be invariant by permutation. Thus, the average of all socially  optimal profiles  is  socially optimal and must be symmetric, i.e. of the form  $\hat D.$
The proof is thus concluded proceeding as in the proof of Theorem \ref{theo1} to prove Equation \eqref{eq WSO}, and recalling Equation \eqref{eq defPoAG}.
\end{proof}
\begin{proof}[{\bf of Proposition  \ref{prop dist}}]
Let us first remark that in the case where $2\delta\beta \bar t \alpha\ge \rho,$ one can check that $\hat D= A$ is a Social Optimum.  This amounts to saying that $C(A,0)$ is optimal and thus allows to conclude.
We now consider the case where $2\delta\beta \bar t \alpha< \rho$.
One can easily check that for every $\bar t>0$ one can find $2\delta\beta \bar t \alpha< \rho\le2\delta\beta \bar t \alpha\exp (\beta \bar t(1-\alpha) \times \min_{i\in\mathcal{N}}\sum_{k\in\mathcal{N}} a_{i,k})$ such that there exists $0<\bar{\kappa}\le\min_{i\in\mathcal{N}}\sum_{k\in\mathcal{N}} a_{i,k}$ satisfying
\begin{equation}2\delta\beta \bar t \alpha\exp ( \beta \bar t (1-\alpha)\bar{\kappa}) = \rho\,.\label{bound}\end{equation}
Let us then recall that for any $\kappa\ge0$, and $k,\ell\in\mathcal{N}$,
\begin{small}
\begin{equation*}
\dfrac{\partial \hat \Pi} {\partial d_{\{k,\ell\}}}(A-C(A,\kappa))= \\
\sum_{i\in\mathcal{N}} \delta \beta \bar t \alpha \exp\left(\beta \bar t (1-\alpha) C(A,\kappa)\right)_{i,k}
 + \sum_{i\in\mathcal{N}} \delta \beta \bar t  \alpha\exp\left(\beta \bar t (1-\alpha) C(A,\kappa)\right)_{i,\ell}  - \rho\,.
 \end{equation*}
\end{small}
Now, it is straightforward to check that, for $\kappa>0$, the largest eigenvalue of  $ \beta \bar t (1-\alpha)C(A,\kappa)$ is $\mu_1:= \beta \bar t (1-\alpha)\kappa.$ According to the Perron-Frobenius theorem, this largest eigenvalue is simple. Furthermore, the associated normalised eigenvector is $ \v= ({1}/{\sqrt{N}})\mathbbm{1}.$  Thus, applying  Lemma \ref{le eigenvec}, one gets
 \begin{equation*} \exp\left(\beta \bar t (1-\alpha) C(A,\kappa)\right)=   \exp (\beta \bar t (1-\alpha) \kappa)  \V\left(1 +\mathcal{O}\left(\exp( -|\mu_1-\mu_2 |) \right)\right)\,,
\end{equation*}
where $\V:=\v \v^\top$ and $\mu_2$ denotes the second largest eigenvalue in module.

One shall then  notice that for all $i,j \in \mathcal{N},$ $(\V)_{i,j}=1/N,$ so that
\begin{small}
\begin{equation*}
 \dfrac{\partial  \hat\Pi} {\partial d_{\{k,\ell\}}}(A-C(A,\kappa))=
 2 \delta \beta \bar t  \alpha \exp (\beta \bar t(1-\alpha)  \kappa) \left(1 +\mathcal{O}\left(\exp( -|\mu_1-\mu_2 |) \right)\right)  - \rho \,.
 \end{equation*}
 \end{small}
 Noting that, all the other parameters being fixed,  the spectral gap $|\mu_1-\mu_2|$ is increasing with respect to $\bar{t}$ and $\kappa,$ one can assume that for every $\varepsilon>0,$ there exists $\bar T>0$ such
that for $\bar t\ge \bar T$, $$2 \delta \beta \bar t  \alpha \exp ( \beta \bar t(1-\alpha) \kappa) \left(1 +\mathcal{O}\left(\exp( -|\mu_1-\mu_2 |) \right)\right)\leq  2 \delta \beta \bar t  \alpha \exp (\beta \bar t(1-\alpha)  \kappa)+\nicefrac{\varepsilon}{\| A\|}\,,$$ for all $\kappa>0.$
  Combining the latter with Equation \eqref{bound}, one concludes that  $C(A,\bar\kappa)$ is an approximate critical point in the sense  that for all $\{k,\ell\} \in \mathcal{E},$
  \begin{equation} \left| \dfrac{\partial  \hat\Pi} {\partial d_{\{k,\ell\}}}(A- C(A,\bar\kappa))\right| \leq \nicefrac{\varepsilon}{\| A\|}\,. \label{norm1} \end{equation}
  Furthermore, if $\hat D$ denotes the Social Optimum, one has  by construction  \begin{equation} \|A- C(A,\bar\kappa)- \hat D \| \leq \| A\| \,. \label{norm2}\end{equation}
    Now, $\hat\Pi$ being continuous and differentiable, one gets through the  mean value theorem
  \begin{equation*}  | \hat\Pi (A-C(A,\bar\kappa))- \hat\Pi (\hat D)   | \leq \left| \dfrac{\partial  \hat\Pi} {\partial d_{\{k,\ell\}}}(A-C(A, \bar\kappa))\right| \times\|A-C(A,\bar\kappa)- \hat D \| \,, \end{equation*}
leading, using Equations \eqref{norm1} and  \eqref{norm2}, to the required result
   \begin{equation*}  | \hat \Pi (A-C(A,\bar\kappa))- \hat\Pi (\hat D)   | \leq  \varepsilon \,. \end{equation*}
\end{proof}
\begin{proof}[{\bf of Proposition \ref{prop exPOK}}]
We assume without loss of generality that $i=1$. By concavity of $\Pi_1$ the set of local optima  is convex. Moreover, given  the asymmetry of the game (involving only player $1$), the set of local optima shall be invariant by the permutation of nodes leaving node $1$ invariant. Thus, the average of all locally optimal profiles is locally optimal.  We let $\bar D$ be such optimum. It is, in particular, such that there exists $h\in[0,a]$ such that for all $1<j\le N,\,\bar d_{1,j}=\bar d_{j,1}=h$. For $1\le j\le N,$ $C_{1,j}(A,\bar D,\beta,\bar t,\x_0)$ is of the form
\begin{align*}\delta_1\beta\bar t\alpha(\exp(\bar H))_{1,j}\,,
\end{align*}
where $\bar H$ writes
\begin{align*}
\bar H\defeq
\left(\begin{matrix}0&\mu&\dots&\dots&\mu\\\mu&\ddots&0&\dots&0\\\vdots&0&\ddots&\ddots&\vdots\\
 \vdots&\vdots&\ddots&\ddots&0\\
\mu&0&\dots&0&0\end{matrix}\right)\,\,\mbox{ with }\, \,\mu:=\beta \bar t(1-\alpha)(a- h)\,.
\end{align*}
Using a Taylor expansion, we obtain that
$$(\exp(\bar H))_{1,1}=\sum_{k\ge 0}\frac{1}{2k!}(N-1)^k\mu^{2k}=\cosh(\sqrt{N-1}\mu)\,,$$
while for $1<j\le N,$ $$(\exp(\bar H))_{1,j}=\sum_{k\ge 0}\frac{1}{(2k+1)!}(N-1)^k\mu^{2k+1}=\frac{\sinh(\sqrt{N-1}\mu)}{\sqrt{N-1}}\,.$$
Hence, in view of the assumption on $\rho$, $\bar D$ is an Homogeneous Interior Equilibrium. Moreover, global strategies dominate local ones if and only if for some $1< j\le N,$ $C_{1,j}(A,\bar D,\beta,\bar t,\x_0)>C_{1,1}(A,\bar D,\beta,\bar t,\x_0)$ and thus if
$$\sinh(\sqrt{N-1}\mu)>\sqrt{N-1}\cosh(\sqrt{N-1}\mu)\,.$$
\end{proof}

\bibliographystyle{apa}
\small{
\bibliography{BibGB}}

\end{document}

%% file: basecom.tex
\renewcommand{\P}{\mathbb{P}}

\newcommand{\R}{\mathbb{R}}

\newcommand{\HP}[1] 
    {\ensuremath{\mathscr{H}^{#1}}}

